\documentclass[amsmath,amssymb,aps,prd,superscriptaddress,floatfix,nofootinbib]{revtex4-2}

\usepackage{graphicx}
\usepackage{dcolumn}
\usepackage{bm}
\usepackage{hyperref}
\usepackage{amsmath}
\usepackage{placeins}
\usepackage{color}

\hypersetup{colorlinks, linkcolor = [rgb]{0,0.0,0.75}, citecolor = [rgb]{0,0.0,0.75}, urlcolor = [rgb]{0,0.0,0.75}}

\newcommand{\bss}[1]{\ensuremath{{\boldsymbol{#1}}}}
\newcommand{\op}{O}
\newcommand{\xb}{\mathbf{x}}

\begin{document}

\title{Binding energy of the \texorpdfstring{$\bm{T_{bb}}$}{Tbb} tetraquark from lattice QCD with relativistic and nonrelativistic heavy-quark actions}

\author{Jakob Hoffmann}
\affiliation{Goethe-Universit\"at Frankfurt am Main, Institut f\"ur Theoretische Physik, Max-von-Laue-Stra{\ss}e 1, D-60438 Frankfurt am Main, Germany}

\author{Stefan Meinel}
\affiliation{Department of Physics, University of Arizona, Tucson, AZ 85721, USA}

\begin{abstract}
We present a new determination of the $\bar b \bar b u d$ ($J^P=1^+$, $I=0$) tetraquark binding energy using lattice QCD with domain-wall light quarks and a nonperturbatively tuned three-parameter anisotropic-clover ``relativistic'' action for the $b$ quarks. We also perform a direct comparison with a reanalysis of data generated in prior work using a lattice-NRQCD action for the $b$ quarks and otherwise identical parameters. Using the new data with relativistic $b$ quarks from seven different ensembles with multiple lattice spacings and pion masses, we perform combined chiral and continuum extrapolations and obtain $(m_{T_{bb}}-m_B-m_{B^*})_{\rm RHQ}=(-76 \pm 23)$ MeV. For the NRQCD data from five ensembles, we perform chiral-only extrapolations and obtain $(m_{T_{bb}}-m_B-m_{B^*})_{\rm NRQCD}=(-74 \pm 17 \pm 10)$ MeV. The lower magnitude of the results obtained here, compared to the original analysis in \href{https://doi.org/10.1103/PhysRevD.100.014503}{Phys.~Rev.~D \textbf{100}, 014503 (2019)}, is due to the use of the symmetric parts of the correlation matrices with local four-quark operators only.
\end{abstract}

\maketitle

\section{Introduction}
\label{sec:intro}

The existence of a QCD-stable $\bar b \bar b u d$ tetraquark meson with quantum numbers $I(J^P) = 0(1^+)$ (in short, $T_{bb}$) is now well-established theoretically, but there remains a considerable uncertainty in the binding energy  \cite{Ader:1981db,Carlson:1987hh,Manohar:1992nd,SilvestreBrac:1993zem,Brink:1998as,Vijande:2003ki,Janc:2004qn,Vijande:2006jf,Navarra:2007yw,Ebert:2007rn,Zhang:2007mu,Lee:2009rt,Bicudo:2012qt,Brown:2012tm,Bicudo:2015kna,Bicudo:2015vta,Bicudo:2016ooe,Francis:2016hui,Karliner:2017qjm,Eichten:2017ffp,Wang:2017uld,Junnarkar:2018twb,Park:2018wjk,Deng:2018kly,Wang:2018atz,Leskovec:2019ioa,Liu:2019stu,Hernandez:2019eox,Hudspith:2020tdf,Mohanta:2020eed,Tan:2020ldi,Lu:2020rog,Braaten:2020nwp,Faustov:2021hjs,Guo:2021yws,Dai:2022ulk,Kim:2022mpa,Chen:2022ros,Praszalowicz:2022sqx,Richard:2022fdc,Wu:2022gie,Maiani:2022qze,Song:2023izj,Liu:2023vrk,Hudspith:2023loy,Aoki:2023nzp,Meng:2023jqk,Feijoo:2023sfe,Ren:2023pip,Alexandrou:2024iwi,Colquhoun:2024jzh,Francis:2024fwf,Li:2023wug,Meng:2024yhu,Berwein:2024ztx,Hoffer:2024fgm,Jiang:2024hxq,Brambilla:2024imu,Zhang:2025wmr,Prelovsek:2025vbr,Nagatsuka:2025szy,Tripathy:2025vao,Vujmilovic:2025czt}. In lattice-QCD calculations of the $T_{bb}$ binding energy \cite{Francis:2016hui,Junnarkar:2018twb,Leskovec:2019ioa,Mohanta:2020eed,Hudspith:2023loy,Alexandrou:2024iwi,Colquhoun:2024jzh,Prelovsek:2025vbr,Nagatsuka:2025szy,Tripathy:2025vao,Vujmilovic:2025czt}, two important sources of systematic errors are (i) heavy-quark discretization errors and (ii) the difficulty to extract the true ground-state energy from fits to correlation functions. In this work, we will compare results from two different heavy-quark actions in a controlled manner to study the former, while aiming to avoid issues with the ground-state energy extraction that have become evident in recent years.

With the exception of Refs.~\cite{Nagatsuka:2025szy,Vujmilovic:2025czt}, all of the published direct lattice studies of the $T_{bb}$ used some variant of lattice NRQCD \cite{Thacker:1990bm,Lepage:1992tx} with tadpole improvement \cite{Lepage:1992xa,Shakespeare:1998dt} for the $b$ quarks. The various calculations differ in the number of operators included in the NRQCD action and in the schemes for setting the matching coefficients. For example, Ref.~\cite{Francis:2016hui} includes the bilinear operators through order $v^4$ in the heavy-heavy power counting with tree-level matching coefficients; Ref.~\cite{Leskovec:2019ioa} uses the same set of operators, but employs one-loop lattice perturbation theory for the coefficient of the $\bss{\sigma}\cdot\mathbf{B}$ term \cite{Hammant:2011bt} and tree level for the remaining operators, while Ref.~\cite{Alexandrou:2024iwi} again includes the same set of operators but with one-loop matching for the ``kinetic'' terms \cite{HPQCD:2011qwj} and tree-level matching for the other operators. Reference \cite{Hudspith:2023loy} instead uses a neural-network-based nonperturbative tuning of the NRQCD matching coefficients based on the bottomonium spectrum, which yields values that differ drastically from the perturbative expectation and leads to underestimated heavy-light hyperfine splittings. It is difficult to disentangle the effects of the various choices on the $T_{bb}$ binding energy, as the aforementioned calculations differ also in other aspects, such as the hadron interpolating operators, the correlation function fit methodologies, and the light-quark and gluon actions. More fundamentally, a disadvantage of lattice NRQCD is that computations can only be performed for $a m_b \gtrsim 1$, and NRQCD matching/truncation errors remain even if a continuum extrapolation is carried out. An alternative approach that allows computations at the physical $b$-quark mass for arbitrary lattice spacings is to use a Wilson-like action with heavy-quark-specific parameter tuning conditions chosen to remove heavy-quark discretization errors from physical observables \cite{El-Khadra:1996wdx,Chen:2000ej,Aoki:2001ra,Aoki:2003dg,Christ:2006us,Lin:2006ur,RBC:2012pds}. This approach, known as the \emph{Fermilab method} or \emph{relativistic heavy quark (RHQ) action}, has already been used to study the $T_{bb}$ in Refs.~\cite{Nagatsuka:2025szy,Vujmilovic:2025czt}. However, the calculation of Ref.~\cite{Nagatsuka:2025szy} did not include local four-quark interpolating operators and therefore likely failed to resolve the true ground state, while Ref.~\cite{Vujmilovic:2025czt} (which focused on the electromagnetic form factors) was limited to a single heavier-than-physical pion mass and a single lattice spacing. Here, we present a new lattice calculation of the $T_{bb}$ binding energy using a nonperturbatively tuned three-parameter
RHQ action for the $b$ quarks and a domain-wall action for the light quarks (both valence and sea quarks). The calculation is performed on seven different ensembles of gauge configurations generated by RBC/UKQCD, with lattice spacings in the range from $0.114$ to $0.073$ fm and pion masses in the range from 431 to 139 MeV, allowing us to perform a combined chiral and continuum extrapolation. Moreover, five of the ensembles, as well as the smearing parameters in the interpolating fields, are shared with the previous NRQCD calculation of Ref.~\cite{Leskovec:2019ioa}, allowing us to perform a controlled comparison between RHQ and NRQCD results. To ensure that the $b$-quark action is the only difference in this comparison, we reanalyze the data of Ref.~\cite{Leskovec:2019ioa} using the same fit methodology as used in the RHQ case. 

Due to the exponential signal-to-noise problem, to reliably determine the $T_{bb}$ binding energy from a lattice calculation with limited computational resources, good choices of interpolating operators and fit methodologies are needed. An analysis of the available direct lattice calculations \cite{Francis:2016hui,Junnarkar:2018twb,Leskovec:2019ioa,Mohanta:2020eed,Meinel:2022lzo,Hudspith:2023loy,Alexandrou:2024iwi,Colquhoun:2024jzh,Parrott:2024rsh,Francis:2024fwf,Prelovsek:2025vbr,Nagatsuka:2025szy,Vujmilovic:2025czt} suggests that calculations with gauge-fixed ``wall'' quark-field sources and point-like sinks may suffer from a substantial negative bias in the $T_{bb}$ energy (and hence overestimate the magnitude of the binding energy) if the fit is performed at insufficient time separation. For two-point functions $C(t)=\sum_n \langle \Omega |O_{\rm sink}(t)|n\rangle\langle n|O_{\rm source}^\dag(0)|\Omega\rangle e^{-E_n t}$ with different operators at source and sink, the product of matrix elements multiplying the exponential function may vary in sign among the different states $n$, which allows the effective energy $E_{\text{eff}}(t)=\frac{1}{a}\text{ln}(C(t)/C(t+a))$ to approach the ground-state energy for large $t$ from either direction or in a non-monotonic way, with the possibility of ``false plateaus'' at intermediate times, especially for multi-quark systems with a dense low-lying finite-volume spectrum (similar issues are well known from lattice studies of two-nucleon systems; see, e.g., Refs.~\cite{Iritani:2016jie,Yamazaki:2017jfh,Horz:2020zvv,Nicholson:2021zwi,Green:2025rel,BaryonScattering:2025ziz,Detmold:2026vjx}). For the $T_{bb}$, it was shown that this problem can be reduced by using extended ``box'' sinks \cite{Hudspith:2020tdf,Colquhoun:2024jzh}. Alternatively, equal-width Gaussian-smeared quark fields can be used in the operators at both source and sink to ensure that $E_{\text{eff}}(t)$ (of diagonal correlation functions or principal correlators obtained through the generalized eigenvalue problem (GEVP) \cite{Blossier:2009kd}) approaches the ground state monotonically from above. Reference \cite{Leskovec:2019ioa} used such Gaussian smearing, but included also $B$-$B^*$ and $B^*$-$B^*$ ``scattering operators,'' in which the $B$ and $B^*$
operators are individually projected to zero momentum. The scattering operators were included at the sink only, so that the entire calculation could be performed with previously generated light-quark propagators with Gaussian-smeared point sources. In combination with three different local four-quark operators, this led to $(5\times 3)$ correlation matrices, to which multi-exponential matrix fits were performed. It was observed that the inclusion of the scattering operators at the sink led to lower extracted ground-state energies, which, at the time, were deemed more reliable. More recently, a computation of the full $(5\times5)$ symmetric correlation matrices with  local and scattering operators at both source and sink was performed \cite{Alexandrou:2024iwi}, and these correlation matrices were analyzed using the GEVP. A comparison of the principal correlators obtained from the full $(5\times 5)$ GEVP with those from a $(3\times 3)$ GEVP using only the local operators showed that, while the inclusion of scattering operators is crucial for correctly determining the excited (two-meson-like) states of the $\bar{b}\bar{b}ud$ system on the lattice, there is no visible difference in the lowest principal correlator used to extract the $\bar{b}\bar{b}ud$ ground-state energy (see Fig.~5 of Ref.~\cite{Alexandrou:2024iwi}). Since we are only interested in the ground-state energy here, we therefore compute and analyze only $(3\times 3)$ correlation matrices with local operators in this work, and reanalyze the corresponding $(3\times 3)$ sub-matrices of Ref.~\cite{Leskovec:2019ioa}, extracting the lowest principal correlator using the GEVP in both cases.

\FloatBarrier
\section{Lattice actions and parameters}

\begin{table}
 \begin{tabular}{lccccccccccccccccccc}
\hline\hline
Ensemble & $N_s^3\times N_t \times N_5$ & $\beta$   & DW action  &  $am_{u,d} $ & $am_{s}$   & $a^{-1}$ [GeV] &  $m_\pi$ [GeV]  & $N_{\rm ex}^{\rm (NR)}$ & $N_{\rm sl}^{\rm (NR)}$  & $N_{\rm ex}^{\rm (R)}$ & $N_{\rm sl}^{\rm (R)}$  \\
\hline
C00078     & $48^3\times96\times24$ & $2.13$  & M\"obius  & $0.00078$   & $0.0362$   & $1.7295(38)$ & $0.13917(35)$  & 80  & 2560  & 158  & 5056 \\
C005LV & $32^3\times64\times16$ & $2.13$  & Shamir  & $0.005$  & $0.04$    & $1.7848(50)$   & $0.3398(12)$    & -- & -- & 186 & 5022 \\
C005   & $24^3\times64\times16$ & $2.13$  & Shamir   & $0.005$   & $0.04$    & $1.7848(50)$   & $0.3398(12)$   & 311 & 9952 & 311 & 9952 \\
C01   & $24^3\times64\times16$ & $2.13$  & Shamir   & $0.01$   & $0.04$    & $1.7848(50)$   & $0.4312(12)$   & 283 & 9056 & 283 & 9056 \\
F004   & $32^3\times64\times16$ & $2.25$  & Shamir    & $0.004$   & $0.03$  & $2.3833(86)$   & $0.3036(14)$    & 251 & 8032  & 251 & 8032 \\
F006   & $32^3\times64\times16$ & $2.25$  & Shamir    & $0.006$    & $0.03$  & $2.3833(86)$   & $0.3607(16)$   & 445 & 14240 & 445 & 14240 \\
F1M    & $48^3\times96\times12$ & $2.31$  & M\"obius & $0.002144$  & $0.02144$ & $2.708(10)$   & $0.2320(10)$    & -- & -- & 226 & 7232 \\
\hline\hline
\end{tabular}
\caption{\label{tab:ensembles}Parameters of the lattice gauge ensembles \cite{RBC:2010qam,RBC:2014ntl,Boyle:2018knm} and light-quark propagators. Here, $N_{\rm ex}^{\rm (NR/R)}$ and $N_{\rm sl}^{\rm (R/NR)}$ denote the number of ``exact'' and ``sloppy'' samples used for the all-mode-averaging procedure \cite{Blum:2012uh,Shintani:2014vja} for the data sets with NRQCD (NR) and relativistic (R) $b$ quarks. The NRQCD data were taken from Ref.~\cite{Leskovec:2019ioa} and are available for five of the seven ensembles only, and with only half the number of samples in the case of C00078.}
\end{table}

\begin{table}
 \begin{tabular}{lccc}
\hline\hline
Ensemble & $a m_Q$ & $\nu$ & $c_E=c_B$ \\
\hline
C00078            & $8.1476$ & $3.3743$ & $5.3944$ \\
C005LV, C005, C01 & $7.3258$ & $3.1918$ & $4.9625$ \\
F004, F006        & $3.2823$ & $2.0600$ & $2.7960$ \\
F1M               & $2.3867$ & $1.8323$ & $2.4262$ \\
\hline\hline
\end{tabular}
\caption{\label{tab:RHQparams}Parameters of the RHQ action used for the $b$ quark \cite{Meinel:2023wyg}.}
\end{table}

We use gauge configurations generated by the RBC and UKQCD collaborations \cite{RBC:2010qam,RBC:2014ntl,Boyle:2018knm} with 2+1 flavors of domain-wall fermions \cite{Kaplan:1992bt,Furman:1994ky,Shamir:1993zy,Brower:2012vk} and the Iwasaki gauge action \cite{Iwasaki:1984cj}; the main parameters of the seven ensembles used here are listed in Table \ref{tab:ensembles}. The same domain-wall action as used for the sea quarks is used for the light valence quarks; this is a Shamir action \cite{Shamir:1993zy} for all ensembles except C00078 and F1M, which use a M\"obius action \cite{Brower:2012vk}. Our calculation uses smeared point-to-all propagators for the up and down quarks (the smearing parameters are given in Sec.~\ref{sec:interpolators}), which were originally generated for Refs.~\cite{Meinel:2020owd,Meinel:2021rbm,Meinel:2021mdj,Meinel:2023wyg}. All-mode-averaging \cite{Blum:2012uh,Shintani:2014vja} over different source positions is used, where on each configuration a small number of samples of ``exact'' correlation functions is combined with a large number of samples of ``sloppy'' correlation functions in such a way that the expectation value is equal to the exact expectation value, but the variance is reduced significantly \cite{Blum:2012uh,Shintani:2014vja}. Low-mode deflation was used for the light-quark conjugate gradient solver, and for the sloppy samples the light-quark-propagator iteration count was fixed to a reduced value.

The RHQ approach used here for the $b$ quark is based on the finding that a Wilson-like lattice action can accurately describe on-shell observables for spatial momenta that are small compared to the inverse lattice spacing even at large quark masses with $m_Q a>1$. An analysis using Symanzik and heavy-quark effective theories shows that discretization errors that would scale as powers of $m_Q a$ can be removed to all orders by giving up the space-time axis-interchange symmetry and appropriately tuning a small number of bare parameters as functions of $m_Q a$ \cite{El-Khadra:1996wdx,Chen:2000ej,Aoki:2001ra,Aoki:2003dg,Christ:2006us,Lin:2006ur,RBC:2012pds}. The action used here has the form
\begin{equation}
 S_Q = a^4 \!\sum_x \bar{Q} \bigg[ {m_Q} + \gamma_0 \nabla_0 - \frac{a}{2} \nabla^{(2)}_0 + {\nu}\!\sum_{i=1}^3\left(\gamma_i \nabla_i - \frac{a}{2} \nabla^{(2)}_i\right)
- {c_E}  \frac{a}{2}\! \sum_{i=1}^3 \sigma_{0i}F_{0i} - {c_B} \frac{a}{4}\! \sum_{i,\, j=1}^3 \! \sigma_{ij}F_{ij}  \bigg] Q.
\end{equation}
Following Refs.~\cite{Christ:2006us,Lin:2006ur,RBC:2012pds}, the three parameters $a m_Q$, $\nu$, and $c_E=c_B$ were tuned nonperturbatively for each lattice spacing such that the $B_s$ meson rest mass $m_{B_s,{\rm rest}}=E_{B_s}(\mathbf{0})$ agrees with experiment, the $B_s$ meson ``speed of light'' squared $c^2=m_{B_s,{\rm rest}}/m_{B_s,{\rm kin}}= [E_{B_s}(\mathbf{p})^2 - E_{B_s}(\mathbf{0})^2]/\mathbf{p}^2$ is consistent with 1, and the hyperfine splitting $E_{B_s^*}(\mathbf{0})-E_{B_s}(\mathbf{0})$ agrees with experiment. The tuning is described in detail in Ref.~\cite{Meinel:2023wyg}, and the parameters are shown again here in Table \ref{tab:RHQparams}. Note that the rapid fall-off of the RHQ $b$-quark propagators with distance means that the standard four-dimensional residual norm is not a reliable criterion for accuracy at large time separations, and an increased solver iteration count must be used.

The lattice NRQCD approach \cite{Thacker:1990bm,Lepage:1992tx} is a direct lattice discretization of continuum NRQCD \cite{Caswell:1985ui}, an effective field theory in which degrees of freedom near and above the heavy-quark mass have been removed, and which corresponds to an expansion in powers of the orbital velocity $v$ of the heavy quarks inside quarkonium or in powers of $\Lambda_{\rm QCD}/m_Q$ when applied to a singly heavy hadron. In contrast to lattice HQET \cite{Eichten:1989zv,Eichten:1989kb,Heitger:2003nj,Sommer:2010ic}, in lattice NRQCD, higher-dimension operators are kept inside the action used for the nonperturbative path integral rather than treated order by order in $\Lambda_{\rm QCD}/m_Q$ as insertions in correlation functions. This is essential for applications to heavy-heavy hadrons, but restricts lattice-NRQCD computations to lattice spacings that satisfy $m_Q a  >1$. 

The form and parameters of the lattice NRQCD action used for the data sets of Ref.~\cite{Leskovec:2019ioa} that we reanalyze here can be found in Refs.~\cite{Brown:2014ena,Leskovec:2019ioa}. This is a tadpole- and Symanzik-improved order-$v^4$ action with the mass parameter tuned nonperturbatively using the spin-averaged bottomonium kinetic mass \cite{Meinel:2010pv} (see Appendix \ref{sec:kineticmasses}), and the matching coefficients computed perturbatively at one-loop level for the chromomagnetic dipole operator \cite{Hammant:2011bt} and at tree level for the other operators.

The kinetic masses of the $B$, $B^*$, $\eta_b$, and $\Upsilon$ mesons obtained with the NRQCD and RHQ actions on some of the ensembles considered here are given in Appendix \ref{sec:kineticmasses}.

\FloatBarrier
\section{Interpolating operators and correlation functions}
\label{sec:interpolators}

As discussed in Sec.~\ref{sec:intro}, here we only use the three
local four-quark operators 
\begin{align}
\label{eq:op_BBast_total_zero} &\op_1(t) = \sum_{\xb} \Big(\bar{b}(x) \gamma_5 d(x)\Big) \Big(\bar{b}(x) \gamma_j u(x)\Big) - (d \leftrightarrow u), \\
\label{eq:op_BastBast_total_zero} &\op_2(t)  = \epsilon_{j k l} \sum_{\xb} \Big(\bar{b}(x) \gamma_k d(x)\Big) \Big(\bar{b}(x) \gamma_l u(x)\Big) - (d \leftrightarrow u), \\
\label{eq:op_Dd_total_zero} &\op_3(t)  = \sum_{\xb} \Big(\epsilon^{a b c} \bar{b}^b(x) \gamma_j C \bar{b}^{c,T}(x)\Big) \Big(\epsilon^{a d e} d^{d,T}(x) C \gamma_5 u^e(x)\Big) - (d \leftrightarrow u), 
\end{align}
where $x=(t,\mathbf{x})$, for the $T_{bb}$ \cite{Leskovec:2019ioa}, and compute the $3\times 3$ correlation matrices
\begin{equation}
C_{jk}(t)=\langle \op_j(t_{\rm src}+t) \op_k^\dag(t_{\rm src}) \rangle.
\end{equation}
To obtain the $B$ and $B^*$ energies, we use the operators
\begin{align}
\op_{B}(t)&= \sum_{\mathbf{x}}\,  \bar{b}(x)\gamma_5 u(x), \\
\op_{B^*}(t)&= \sum_{\mathbf{x}}\,  \bar{b}(x)\gamma_j u(x), \label{eq:Bstar}
\end{align}
and compute the correlation functions $C_{B^{(*)}}(t)=\langle \op_{B^{(*)}}(t_{\rm src}+t) \op_{B^{(*)}}^\dag(t_{\rm src})\rangle$. For both $C_{jk}(t)$ and $C_{B^{(*)}}(t)$, one of the sums over the spatial position (the one at the source) is removed using translational symmetry. We average $C_{B^{(*)}}(t)$ over forward and backward propagating two-point functions. All quark fields in Eqs.~(\ref{eq:op_BBast_total_zero}-\ref{eq:Bstar}) are smeared using gauge-covariant Gaussian spatial smearing, with spatially APE-smeared gauge links for the light quarks and unsmeared gauge links for the heavy quarks, using the parameters given in Table \ref{tab:smearingparams}.

\begin{table}
	\begin{tabular}{lccccccc} \hline \hline 
		Ensemble   & \multicolumn{4}{c}{Up and down quarks} & \hspace{2ex} & \multicolumn{2}{c}{Bottom quarks} \\
               & $N_\textrm{Gauss}$ & $\sigma_\textrm{Gauss}$ & $N_\textrm{APE}$ & $\alpha_\textrm{APE}$ && $N_\textrm{Gauss}$ & $\sigma_\textrm{Gauss}$  \\ \hline
    C00078     & $100$           & $7.171$           & $25$ & $2.5$ && $10$ & $2.0$ \\ 
    C005LV, C005, C01  & $\phantom{0}30$ & $4.350$ & $25$ & $2.5$ && $10$ & $2.0$ \\
    F004, F006 & $\phantom{0}60$ & $5.728$           & $25$ & $2.5$ && $10$ & $2.0$ \\
    F1M & $130$ & $8.9$           & $25$ & $2.5$ && $10$ & $2.0$ \\
    \hline \hline
	\end{tabular}
	\caption{\label{tab:smearingparams}Parameters for the smearing of the quark fields in the interpolating operators (see Ref.~\cite{Leskovec:2019ioa} for the definitions).}
\end{table}

\FloatBarrier
\section{Analysis of the correlation functions}

We extract the $B$, $B^*$ and $\bar{b}\bar{b}ud$ ground-state energies on each ensemble from single-exponential fits to $C_{B^{(*)}}(t)$ (Sec.~\ref{sec:BBstmasses}) and to the lowest principal correlator $\lambda_0(t;t_0)$ obtained by solving the GEVP for $C_{jk}(t)$ (Sec.~\ref{sec:Tbbenergies}). For a robust extraction of these energies, we use the uniform strategy outlined in the following. First, we perform correlated $\chi^2$-minimizing fits for multiple temporal fit ranges $[t_{\text{min}}/a;t_{\text{max}}/a]$. All fits to the correlators were performed using the \textit{lsqfit} package \cite{lsqfit}. Statistical uncertainties are determined via the bootstrap resampling procedure by repeating the fits for $500$ bootstrap samples. For a given fit range $[t_{\text{min}}/a;t_{\text{max}}/a]$, we take the bootstrap mean of this distribution as the ground-state energy and the bootstrap standard deviation as the corresponding statistical uncertainty. The result of this procedure is a set of energies $\{E_0\}$ with statistical uncertainties and their associated $\chi^2$ values, corresponding to the different fit ranges. To combine the results from different fit ranges into a single representative energy, we employ the Bayesian model-averaging technique proposed in Ref.~\cite{Jay:2020jkz}. Given a set of models $m\in\{1,...,M\}$, their associated energies $\{E_{0,m}\}$, statistical uncertainties $\{\sigma_{E_0,m}\}$ and $\chi^2$ values $\{\chi^2_m\}$, the model-averaged energy $\left<E_0\right>$ and its total uncertainty $\sigma_{E_0}$ are computed as  \cite{Jay:2020jkz}
\begin{align}
\label{eq:model-average}
\left<E_0\right>&=\sum_{m=1}^{M}\, E_{0,m}\, \text{pr}(m|D) , \\
\sigma_{E_0}^2&=\sum_{m=1}^{M}\, \sigma_{E_0,m}^2\, \text{pr}(m|D)+\sum_{m=1}^{M}\, E_{0,m}^2\, \text{pr}(m|D)-\left<E_0\right>^2 ,
\label{eq:model-error}
\end{align}
where $\text{pr}(m|D)$ are normalized model weights given by 
\begin{equation}
\text{pr}(m|D)=C\,\text{exp}\left[-\frac{1}{2}\left(\chi_m^2+2k_m+2N_{\text{cut},m} \right) \right] . \label{eq:modelweights}
\end{equation}
Here, $k_m$ denotes the number of fit parameters and $N_{\text{cut},m}$ denotes the number of data points excluded from the specific fit $m$. The constant $C$ is a normalization factor that ensures that the weights are properly normalized, i.e., $\sum_{m=1}^{M}\, \text{pr}(m|D)$ = 1. In the fits considered in this section, different fit ranges $[t_{\text{min}}/a;t_{\text{max}}/a]$ correspond to different models $m$. The last two terms in Eq.~(\ref{eq:model-error}) account for the systematic uncertainty due to the shifts in $E_0$ when changing the fit range.

\subsection{$B$ and $B^*$ meson energies}
\label{sec:BBstmasses}
\begin{figure}[h]
    \centering
    \begin{minipage}{0.45\textwidth}
        \centering
        \includegraphics[width=\linewidth]{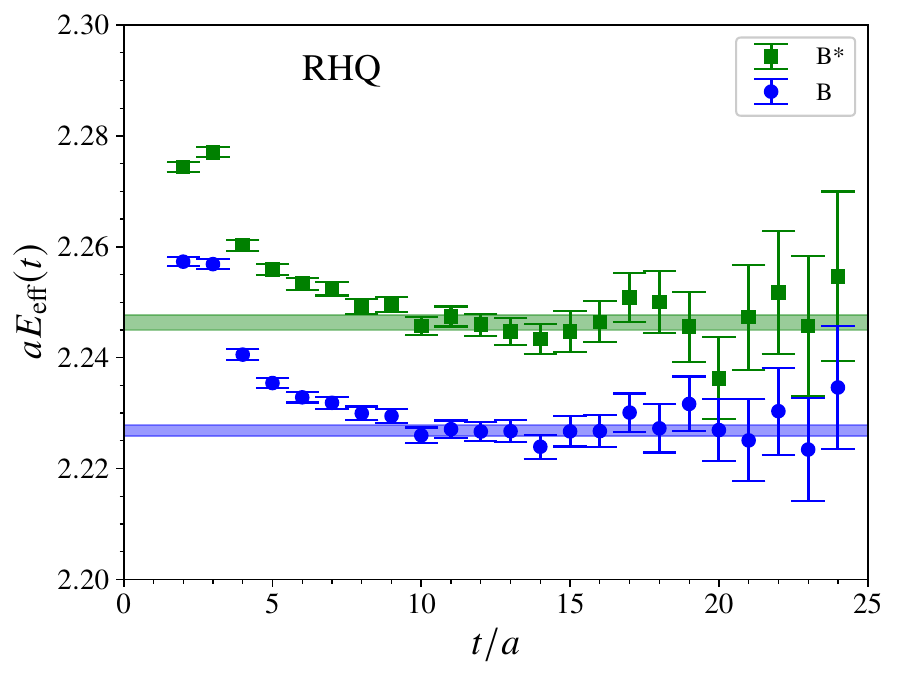}
    \end{minipage}
    \hfill
    \begin{minipage}{0.45\textwidth}
        \centering
        \includegraphics[width=\linewidth]{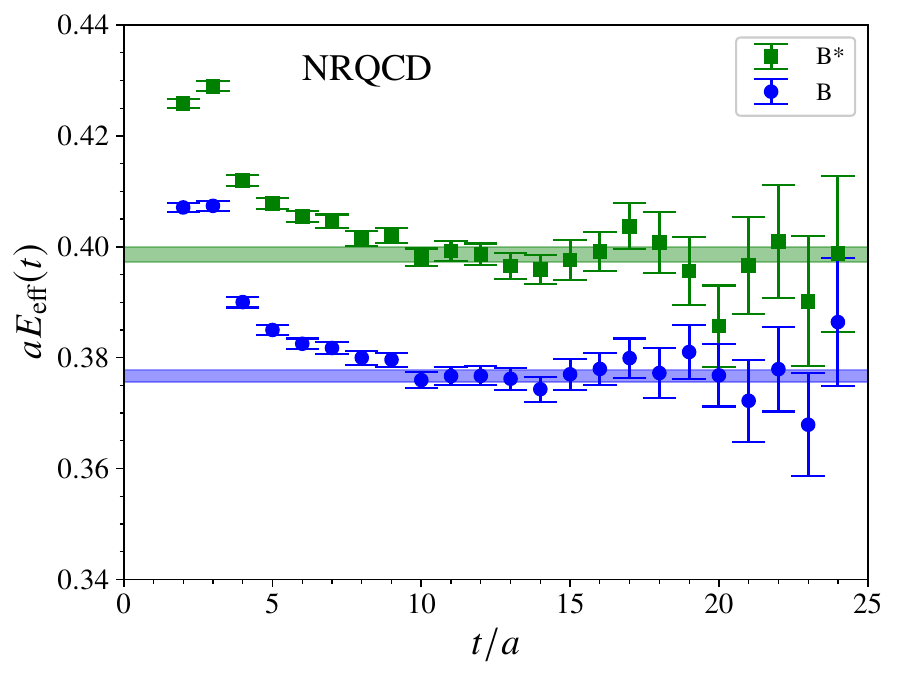}
    \end{minipage}
    \caption{\label{fig:Bmeson-effective-masses}Effective energies of the $B$ and $B^*$ two-point functions from the F006 ensemble for the RHQ action (left) and the NRQCD action (right). The blue and green bands show the model-averaged $B$ and $B^*$ meson energies.}
\end{figure}

To extract the $B$ and $B^*$ meson energies at zero momentum\footnote{With the RHQ action, these energies are equal to the masses. However, with the NRQCD action, all energies are shifted by approximately $-n_b m_b$, where $n_b$ is the number of $b$ quarks inside the hadron, and this shift cancels in appropriate energy differences such as $E_{T_{bb}}-E_B-E_{B^*}$.}, we performed correlated single-exponential fits to $B$ and $B^*$ meson correlation functions computed with the RHQ action on the ensembles listed in Table \ref{tab:ensembles}. Moreover, we also extracted the NRQCD $B$ and $B^*$ meson energies using the correlation functions previously computed in Ref.~\cite{Leskovec:2019ioa}. For the fits, we fixed $t_{\text{max}}/a$ as the largest values at which the signal is not completely destroyed by statistical noise.
The chosen values of $t_{\text{max}}/a$, the ranges for $t_{\text{min}}/a$ used in the model averaging, and the resulting 
estimates for the $B$ and $B^*$ meson energies in lattice units are given in  Table \ref{tab:BBstmasses}. Figure \ref{fig:Bmeson-effective-masses} displays the effective energies $aE_{\text{eff}}(t)=\text{ln}(C(t)/C(t+a))$ for one of the ensembles along with the model-averaged estimates for the $B$ and $B^*$ meson energies.

While not needed to determine the $T_{bb}$ binding energy, it is also interesting to study the hyperfine splittings $\Delta E_{\rm HFS}=E_{B^*}-E_B$ and compare the NRQCD and RHQ results for them. Using the model weights for the $B$ and $B^*$ correlator fits, we computed the model-averaged $\Delta E_{\rm HFS}$ from the model-averaged $B$ and $B^*$ meson energies on all bootstrap samples. The means of $\Delta E_{\rm HFS}$ over all bootstrap samples, converted to physical units, are shown in Table \ref{tab:hyperfine-splittings}. We computed the corresponding uncertainties as the bootstrap standard deviation of the statistical distribution of $\Delta E_{\rm HFS}$, and added the uncertainty of the lattice spacing in quadrature.

We observe that the hyperfine splittings computed with our NRQCD $b$-quark action (with one-loop values for the coefficient $c_4$ of the chromomagnetic dipole operator) are typically $\sim 10\%$ higher than those computed with the RHQ $b$-quark action (the C00078 ensemble is an exception, but the corresponding result has a large statistical uncertainty). The NRQCD results are affected by missing higher-order corrections to $c_4$, which introduce an error of order $\alpha_s^2 \Lambda_{\rm QCD}^2/m_b\sim 2$ MeV, and by missing heavy-light four-quark operators, whose effect could have similar magnitude \cite{Blok:1996iz}.

\begin{figure}
\includegraphics[width=0.5\linewidth]{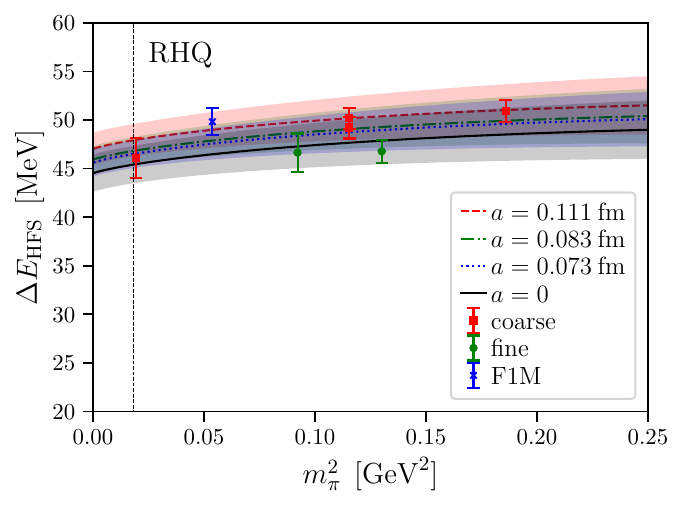}
\caption{\label{fig:HFSRHQ}Chiral-continuum extrapolation of the $B$-meson hyperfine splitting computed using the RHQ action. The curves show the fit model (\ref{eq:HFSchiralcontinuum}) evaluated for the three different lattice spacings for which we have data, and in the continuum limit. The dashed vertical line indicates the physical pion mass.}
\end{figure}

A chiral-continuum extrapolation of the RHQ results for the hyperfine splitting using the prediction from SU(2) heavy-meson chiral perturbation theory \cite{Jenkins:1992hx} with an added $a^2$ term,
\begin{equation}
\Delta E_{\rm HFS}= \Delta E_{\rm HFS,0} - \Delta E_{\rm HFS,0}\frac{3 g^2}{16\pi^2 f^2} m_\pi^2 \ln(m_\pi^2/\mu^2) + c \,m_\pi^2 + d\, a^2, \label{eq:HFSchiralcontinuum}
\end{equation}
with the pion decay constant $f=130.5$ MeV \cite{FlavourLatticeAveragingGroupFLAG:2024oxs} and fit parameters $\Delta E_{\rm HFS,0}$, $g$, $c$, $d$ and Gaussian prior $g=0.5\pm 0.1$ for the axial coupling based on Refs.~\cite{Detmold:2011bp,Bernardoni:2014kla,Flynn:2015xna,Gerardin:2021jch} yields the extrapolated value
\begin{equation}
(\Delta E_{\rm HFS})_{\rm RHQ}=45.4(1.9)\text{ MeV}
\end{equation}
at $m_\pi=135$ MeV and $a=0$, in agreement with the experimental value of 45.18(20) MeV \cite{ParticleDataGroup:2024cfk}. The fit is shown in Fig.~\ref{fig:HFSRHQ}.

\begin{table}[h]
\centering
\renewcommand{\arraystretch}{1.2}
 \begin{tabular}{lccccccc}
\hline\hline
Label  & $(t_{\text{min}}/a)_{\text{RHQ}}$ &$(t_{\text{min}}/a)_{\text{NRQCD}}$ & $t_{\text{max}}/a$ & $(a E_B)_{\text{RHQ}}$ & $(a E_{B^*})_{\text{RHQ}}$ & $(a E_B)_{\text{NRQCD}}$ & $(a E_{B^*})_{\text{NRQCD}}$ \\
\hline
C00078      & $4...8$ & $4...10$& $15$ & $3.0576 (34)$ & $3.0842 (35)$ & $0.4633 (57)$ & $0.4852 (77)$ \\
C005LV & $5...10$ & $5...10$ & $18$ & $2.9729 (13)$ & $3.0010 (15)$ & -- & -- \\
C005   & $5...10$ & $5...10$ & $18$ & $2.9710 (13)$ & $2.9986 (16)$ & $0.4644 (13)$ & $0.4941 (14)$  \\
C01    & $5...10$ & $5...10$ & $20$ & $2.9805 (15)$ & $3.0090 (24)$ & $0.4736 (16)$ & $0.5059 (19)$ \\
F004    & $8...15$ & $8...15$ & $22$ & $2.2252 (16)$ & $2.2448 (24)$ & $0.3760 (18)$ & $0.3962 (30)$ \\
F006    & $8...15$ & $8...15$ & $22$ & $2.2268 (10)$ & $2.2465 (13)$ & $0.3767 (10)$ & $0.3986 (13)$ \\
F1M     & $6...14$ & $6...14$ & $17$  & $1.9541 (14)$ & $1.9725 (16)$ & -- & --\\
\hline\hline
\end{tabular}
\caption{\label{tab:BBstmasses}Model-averaged $B$ and $B^*$ energies in lattice units for all ensembles for the RHQ and NRQCD actions, along with the fit ranges entering in the model averages. Machine-readable files containing the energies are provided in the Supplemental Material \cite{Supplemental}.}
\end{table}

\begin{table}[h]
\centering
\renewcommand{\arraystretch}{1.2}
 \begin{tabular}{lcc}
\hline\hline
Label  & $(\Delta E_{\rm HFS})_{\text{RHQ}}\, [\text{MeV}]$ & $(\Delta E_{\rm HFS})_{\text{NRQCD}}\, [\text{MeV}]$ \\
\hline
C00078    & $46.1 (2.1)$ & $37.7 (4.9)$\\
C005LV & $50.2 (1.1)$ & --\\
C005   & $49.2 (1.1)$ & $53.1 (1.2)$  \\
C01    & $50.9 (1.1)$ & $57.6 (1.3)$\\
F004   & $46.7 (2.0)$ & $48.1 (2.6)$ \\
F006   & $46.8 (1.2)$ & $52.2 (1.5)$  \\
F1M    & $49.8 (1.4)$ & --  \\
\hline\hline
\end{tabular}
\caption{\label{tab:hyperfine-splittings}Model-averaged $B^*-B$ hyperfine splittings for all ensembles for the RHQ and NRQCD actions. Machine-readable files containing these values are provided in the Supplemental Material \cite{Supplemental}.}
\end{table}

\FloatBarrier
\subsection{$T_{bb}$ binding energies}
\label{sec:Tbbenergies}

To extract the $\bar{b}\bar{b}ud$ energy levels from the $3\times 3$ correlation matrices $C_{jk}(t)$ discussed in Sec.~\ref{sec:interpolators}, we solve the generalized eigenvalue problem (GEVP)
\begin{equation}
\sum_{k}\,C_{jk}(t)\,v_{k,n}(t;t_0)=\sum_k\lambda_n(t;t_0)\,C_{jk}(t_0)\,v_{k,n}(t;t_0) 
\end{equation}
with principal correlators $\lambda_n(t;t_0)$ and eigenvectors $v_{k,n}(t;t_0)$, where $t_0$ is a fixed diagonalization time. The energy levels $E_n$ can be extracted from the principal correlators $\lambda_n(t;t_0)$ via correlated single-exponential fits of the form
\begin{equation}
\lambda_n(t;t_0)=A_n\, e^{-E_n(t-t_0)},
\end{equation}
where the prefactors $A_n$ are additional fit parameters $\approx 1$. In the NRQCD case, we used the $3\times 3$ sub-blocks of the full $5\times 3$ correlation matrices computed in Ref.~\cite{Leskovec:2019ioa}; these sub-blocks correspond to the same local for-quark operators as used in the RHQ case. The values of $t_0/a$ were chosen individually for each ensemble. Our strategy was to choose $t_0/a$ at relatively early times for which the signal is well preserved. We scaled the values of $t_0/a$ between ensembles such that $t_0$ in physical units remains approximately constant. In Table \ref{tab:t0-values}, we list the values of $t_0/a$ chosen for each ensemble. We also varied $t_0/a$ in small ranges but did not observe any significant change in the resulting energy levels.

\begin{table}[h]
\centering
\renewcommand{\arraystretch}{1.2}
 \begin{tabular}{lcc}
\hline\hline
Label & $(t_0/a)_{\text{RHQ}}$ & $(t_0/a)_{\text{NRQCD}}$  \\
\hline
C00078     & $3$ & $3$  \\
C005LV & $3$ & -- \\
C005   & $3$ & $3$   \\
C01   & $3$ & $3$   \\
F004   & $4$ & $4$ \\
F006   & $4$ & $4$   \\
F1M    & $4$ & --    \\
\hline\hline
\end{tabular}
\caption{\label{tab:t0-values} The values of the GEVP parameter $t_0/a$ used for each ensemble.}
\end{table}

\begin{figure}[h]
    \centering
    \begin{minipage}{0.45\textwidth}
        \centering
        \includegraphics[width=\linewidth]{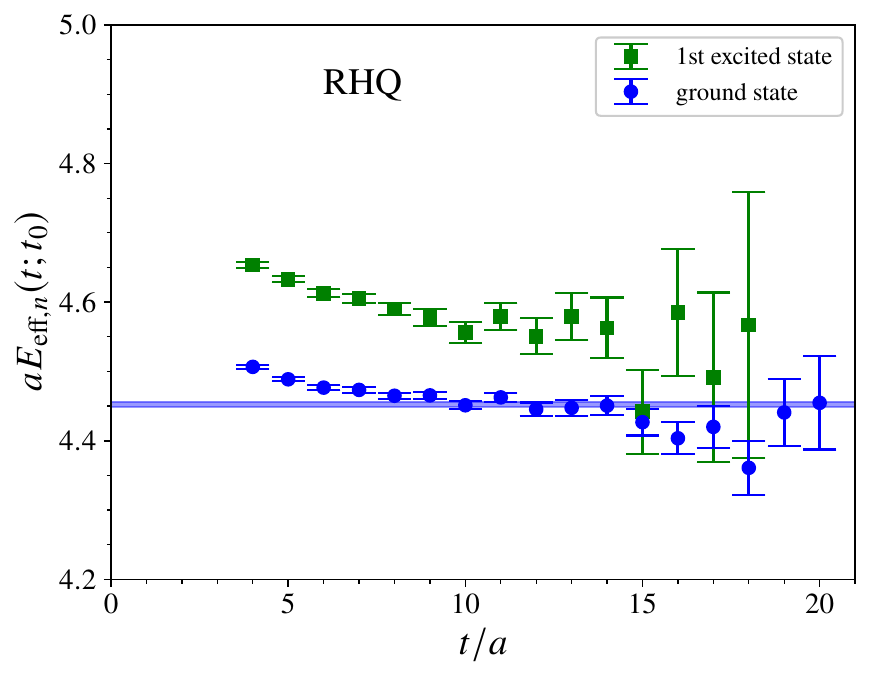}
    \end{minipage}
    \hfill
    \begin{minipage}{0.45\textwidth}
        \centering
        \includegraphics[width=\linewidth]{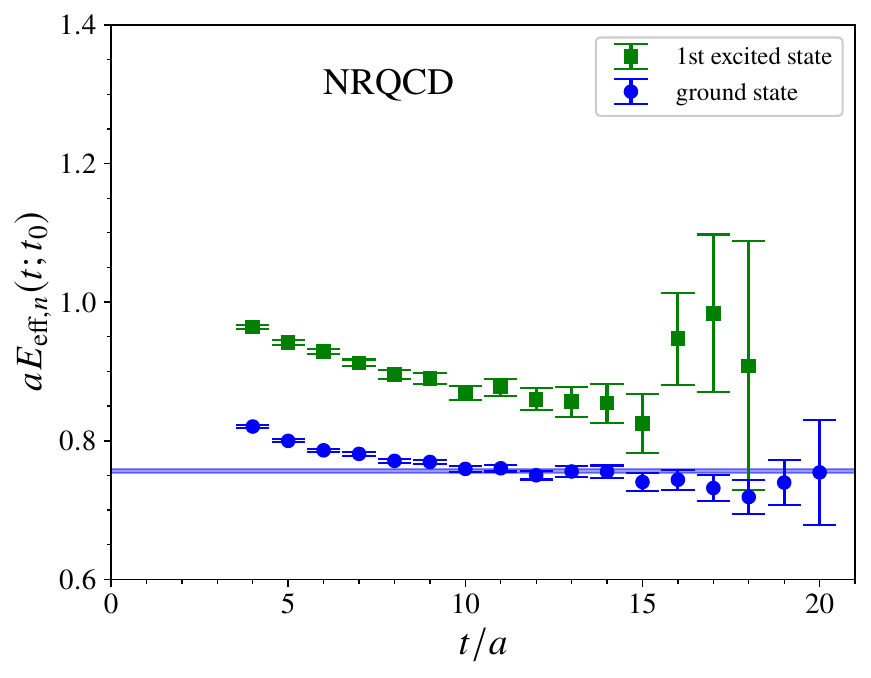}
    \end{minipage}
    \caption{\label{fig:effective-energies-bbud}Effective energies of the $\bar{b}\bar{b}ud$ [$I(J^P)=0(1^+)$] lowest two principal correlators, from the F006 ensemble computed with the RHQ action (left) and the NRQCD action (right). The blue horizontal band represents the model-averaged ground-state energy. Corresponding plots from the other ensembles are shown in Figs.~\protect\ref{fig:EffNRQCD} and \protect\ref{fig:EffRHQ}.}
    
\end{figure}

Figure \ref{fig:effective-energies-bbud} shows the $\bar{b}\bar{b}ud$ effective energies $aE_{\text{eff},n}(t)=\ln(\lambda_n(t;t_0)/\lambda_n(t+a;t_0))$ for $n=0$ and $n=1$. We performed fits only to $\lambda_0(t;t_0)$ and did not attempt to extract excited-state energy levels (which are two-meson-like states), because the omission of the scattering interpolating operators means that the results for these levels are unreliable \cite{Alexandrou:2024iwi}.

The model-averaged results for the $\bar{b}\bar{b}ud$ ground-state energies in lattice units, together with the values of $t_{\text{max}}/a$ and the ranges of 
$t_{\text{min}}/a$ used in the model averaging, are listed in Table \ref{tab:bbudenergies}.

\begin{table}[h]
\centering
\renewcommand{\arraystretch}{1.2}
 \begin{tabular}{lccccc}
\hline\hline
Label  & $(t_{\text{min}}/a)_{\text{RHQ}}$ &$(t_{\text{min}}/a)_{\text{NRQCD}}$ & $t_{\text{max}}/a$ & $(a E_0)_{\text{RHQ}}$  & $(a E_0)_{\text{NRQCD}}$ \\
\hline
C00078      & $4...8$ & $4...10$& $15$ & $6.0596 (81)$ & $0.9007 (83)$ \\
C005LV & $5...10$ & $5...10$ & $18$ & $5.9295 (61)$ & -- \\
C005   & $5...10$ & $5...10$ & $18$ & $5.9375 (39)$ & $0.9248 (34)$  \\
C01    & $5...10$ & $5...10$ & $20$ & $5.9670 (33)$ & $0.9579 (31)$\\
F004    & $8...15$ & $8...15$ & $22$ & $4.4486 (42)$ & $0.7561 (30)$ \\
F006    & $8...15$ & $8...15$ & $22$ & $4.4527 (39)$ & $0.7564 (29)$\\
F1M     & $6...14$ & $6...14$ & $17$ & $3.9004 (43)$ & -- \\
\hline\hline
\end{tabular}
\caption{\label{tab:bbudenergies}Model-averaged  $\bar{b}\bar{b}ud$ ground-state energies in lattice units for all ensembles for the RHQ and NRQCD actions, along with the fit ranges entering in the model averages.  Machine-readable files containing the energies are provided in the Supplemental Material \cite{Supplemental}.}
\end{table}

\begin{figure}[h]
    \centering
    \begin{minipage}{0.47\textwidth}
        \centering
        \includegraphics[width=\linewidth]{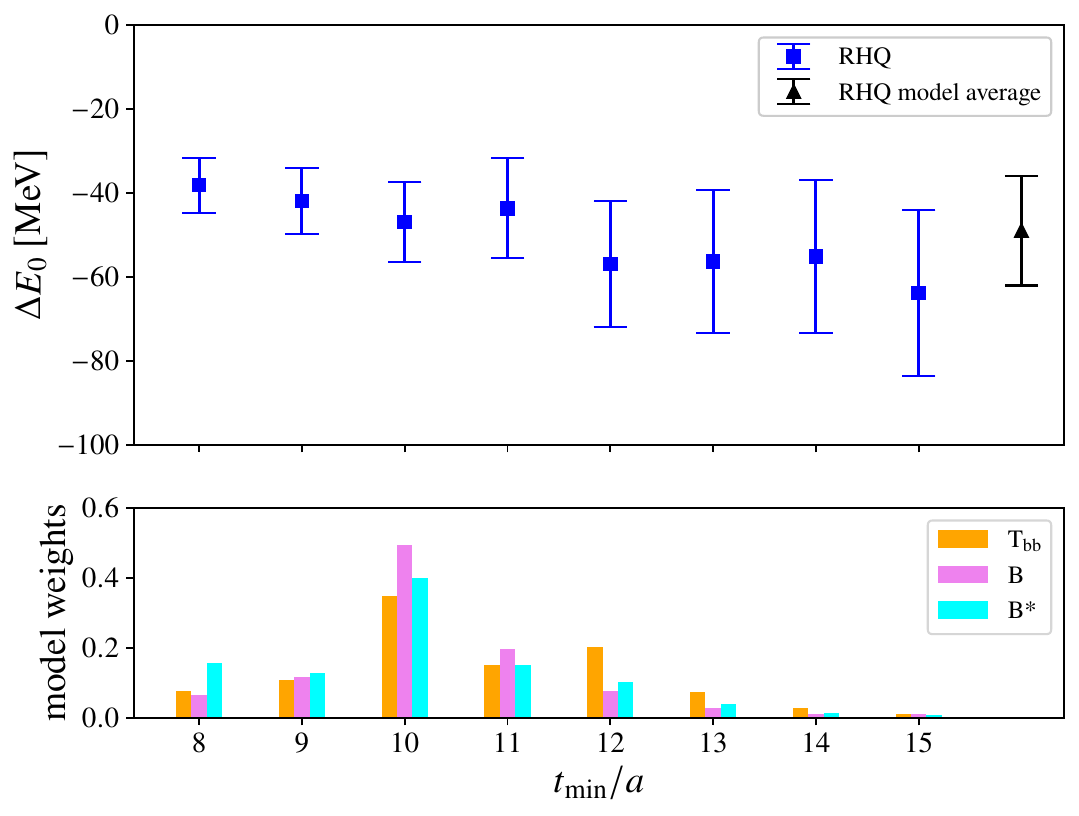}
    \end{minipage}
    \hfill
    \begin{minipage}{0.47\textwidth}
        \centering
        \includegraphics[width=\linewidth]{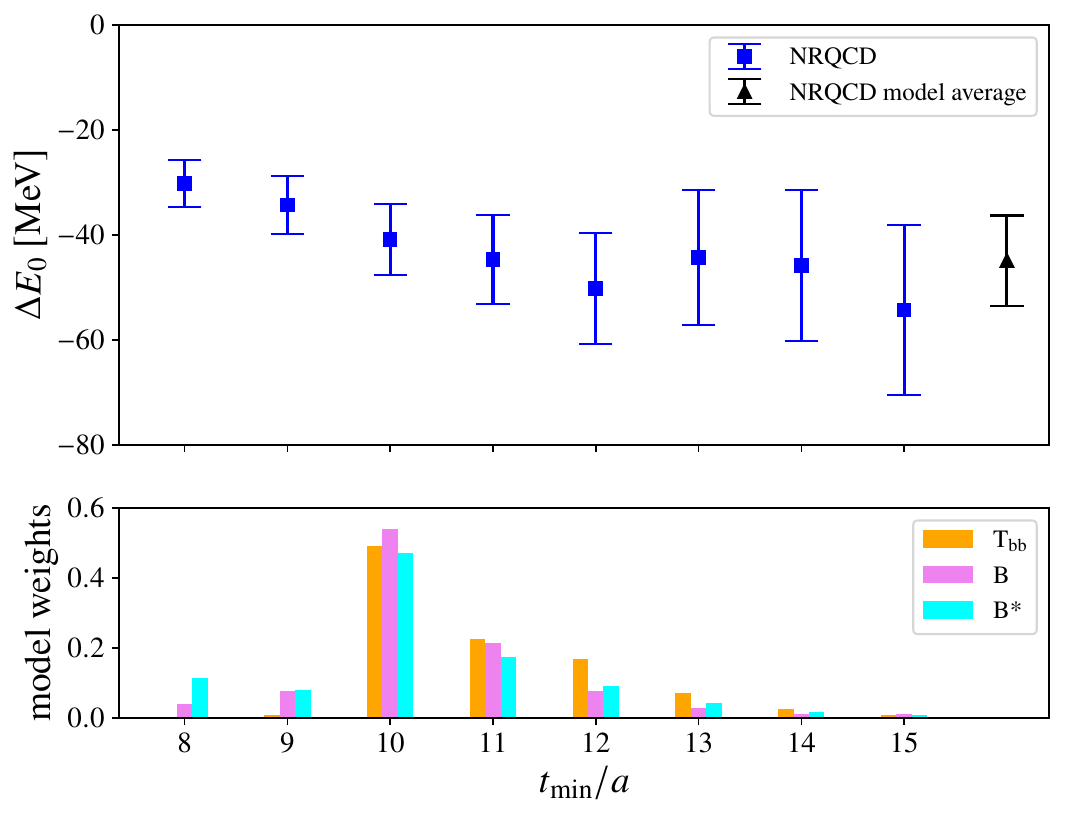}
    \end{minipage}
    \caption{Upper panels: the $T_{bb}$ binding energies $\Delta E_0$ as a function of the value of $t_{\text{min}}/a$ used in the fit of the $T_{bb}$ lowest principal correlator (computed using the model-averaged $B$ and $B^*$ energies), along with the binding energy computed using the model-averaged $T_{bb}$, $B$, and $B^*$ energies. Lower panels: the normalized model weights for the fits to the $T_{bb}$ lowest principal correlator and for the fits to the $B$ and $B^*$ two-point functions, as a function of $t_{\text{min}}/a$. The left and right panels show the results from the RHQ and NRQCD actions, respectively. All results shown here are from the F006 ensemble. Corresponding plots for the other ensembles are shown in Figs.~\protect\ref{fig:DeltaEtminRHQ} and \protect\ref{fig:DeltaEtminNRQCD}.}
    \label{fig:model-averaged-binding-energy}
\end{figure}

Using these results, we computed the binding energies $\Delta E_0=E_0-E_B-E_{B^{*}}$ relative to the $BB^*$ threshold for all ensembles. The statistical uncertainties were propagated as follows. We solved the GEVP and performed  fits on all 500 bootstrap samples for the $t_{\text{min}}/a$ ranges listed in Table \ref{tab:bbudenergies}. For each bootstrap sample, we computed the model averages of $E_0$, $E_B$, and $E_{B^{*}}$ via Eq.~\eqref{eq:model-average} using fixed model weights, which were taken from fits to the original data. We then computed the difference $\Delta E_0=E_0-E_B-E_{B^{*}}$ of the model-averaged quantities for each bootstrap sample. We quote the mean of the bootstrap distribution as the final result for the binding energy and the bootstrap standard deviation as its statistical uncertainty. An illustration of the model averaging is shown in Fig.~\ref{fig:model-averaged-binding-energy} for one of the ensembles. We separately computed the systematic uncertainties in $E_0$, $E_B$, and $E_{B^*}$ due to the variations of the fit ranges using the last two terms of Eq.~\eqref{eq:model-error}. We then added these systematic uncertainties in quadrature to obtain a rough estimate of the systematic uncertainty in $\Delta E_0$. This approach neglects any possible correlations in the systematic uncertainties, likely leading to an overestimate. As such, the computed systematic uncertainty should be regarded as a conservative upper bound rather than a precise value. In Table \ref{tab:binding-energies} we list the final results for the model-averaged binding energies in physical units, where we combined the statistical and systematic uncertainties in quadrature.
As expected, we find that the $T_{bb}$ ground-state energies on all ensembles for both the RHQ and the NRQCD actions lie significantly below the $BB^{*}$ threshold, indicating a deeply bound state that is stable under the strong interaction.

\begin{table}[htbp]
\centering
\renewcommand{\arraystretch}{1.2}
 \begin{tabular}{lcc}
\hline\hline
Label &  $\Delta E_{0,\text{RHQ}}\, [\text{MeV}]$ & $\Delta E_{0,\text{NRQCD}}\, [\text{MeV}]$  \\
\hline
C00078  & $-142 (28)$ & $-84 (22)$  \\
C005LV  & $-79 (13)$ & -- \\
C005    & $-57 (9)$ & $-60 (7)$   \\
C01    & $-40 (8)$ & $-39 (7)$   \\
F004   & $-51 (13)$ & $-38 (11)$ \\
F006    & $-49 (13)$ & $-45 (9)$   \\
F1M    & $-71 (15)$ & --    \\
\hline\hline
\end{tabular}
\caption{\label{tab:binding-energies}Model-averaged $T_{bb}$ binding energies on all ensembles for the RHQ and NRQCD actions. Machine-readable files containing these values are provided in the Supplemental Material \cite{Supplemental}.}
\end{table}

\FloatBarrier
\section{Chiral and continuum extrapolations of the binding energy}
\FloatBarrier

To extract the $T_{bb}$ binding energy at physical point, we perform  chiral and combined chiral-continuum extrapolations of the results presented in Table \ref{tab:hyperfine-splittings}. Following Ref.~\cite{Alexandrou:2024iwi}, we do not attempt to apply the L\"uscher quantization condition to remove finite-volume effects here (the ground state is located below the onset of the left-hand cut in the scattering amplitude caused by one-pion exchange), but we expect finite-volume effects to be negligible compared to our statistical uncertainties, given the deep binding of the $T_{bb}$ and the large values of $m_\pi L$ of the ensembles used.

We consider chiral-only extrapolations of the forms
\begin{align}
\text{``linear'':  }\Delta E_0(m_\pi)&=\Delta E_0(0)+c\,m_\pi^2 , \\
\text{``linear+3/2'':  }\Delta E_0(m_\pi)&=\Delta E_0(0)+c\,m_\pi^2+\widetilde{c}\,m_\pi^3, \\
\text{``linear+log'':  }\Delta E_0(m_\pi)&=\Delta E_0(0)+c\,m_\pi^2 + l\, m_\pi^2 \ln(m_\pi^2/\mu^2) ,
\end{align}
as well as chiral-continuum extrapolations of the forms
\begin{align}
\text{``linear'':  }\Delta E_0(m_\pi,a)&=\Delta E_0(0)+c\,m_\pi^2+d\,a^2 , \\
\text{``linear+3/2'':  }\Delta E_0(m_\pi,a)&=\Delta E_0(0)+c\,m_\pi^2+\widetilde{c}\,m_\pi^3+d\,a^2 , \\
\text{``linear+log'':  }\Delta E_0(m_\pi,a)&=\Delta E_0(0)+c\,m_\pi^2 + l\, m_\pi^2 \ln(m_\pi^2/\mu^2) +d\,a^2.
\end{align}
The latter are applied only to the RHQ data. Above, $\Delta E_0(0)$, $c$, $\widetilde{c}$, $l$, and $d$ are fit parameters, and $\mu$ is an arbitrary renormalization scale. The binding energy is independent of $\mu$, because the $\mu$-dependence of the $m_\pi^2 \ln(m_\pi^2/\mu^2)$ term is canceled by the $\mu$-dependence of the coefficient $c$. The physical-point binding energies obtained using these fits, together with the values of $\chi^2/{\rm d.o.f}$, are listed in Table \ref{tab:extrapolations}.

\begin{table}
\begin{tabular}{llcc}
\hline\hline
Action & Model & \hspace{1.5ex} $\chi^2/{\rm d.o.f.}$ \hspace{1.5ex} & $\Delta E_0(m_{\pi,\text{phys}},(a=0))$ [MeV] \\
\hline
NRQCD  & chiral only, linear         & 1.60 & $-71\pm12$ \\
NRQCD  & chiral only, linear+3/2     & 2.35 & $-77\pm23$ \\
NRQCD  & chiral only, linear+log     & 2.33 & $-78\pm22$ \\
RHQ  & chiral only, linear           & 1.61 & $-90\pm12$ \\
RHQ  & chiral only, linear+3/2       & 1.65 & $-113\pm20$ \\
RHQ  & chiral only, linear+log       & 1.56 & $-117\pm23$ \\
RHQ  & chiral-continuum, linear      & 0.98 & $-64\pm17$ \\
RHQ  & chiral-continuum, linear+3/2  & 0.97 & $-84\pm26$ \\
RHQ  & chiral-continuum, linear+log  & 0.93 & $-88\pm28$ \\
\hline\hline
\end{tabular}
\caption{\label{tab:extrapolations} Physical-point $T_{bb}$ binding energies obtained from chiral and chiral+continuum extrapolations using the different models discussed in the main text, together with the corresponding $\chi^2/{\rm d.o.f.}$ values.}
\end{table}

\begin{figure}
\includegraphics[width=0.49\linewidth]{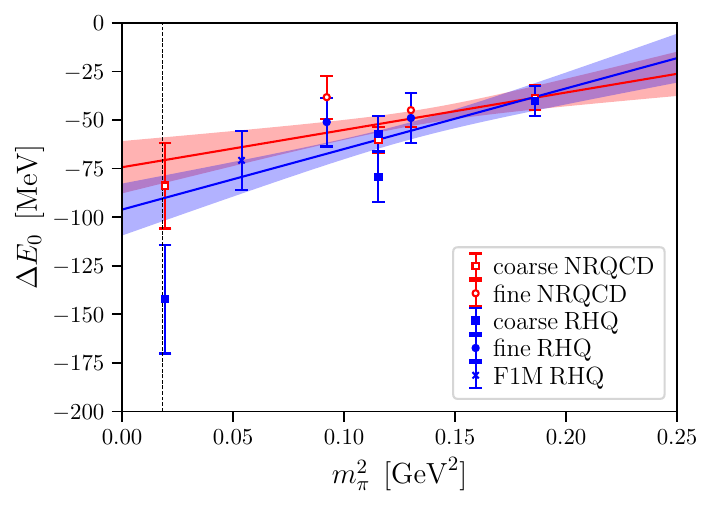} \includegraphics[width=0.49\linewidth]{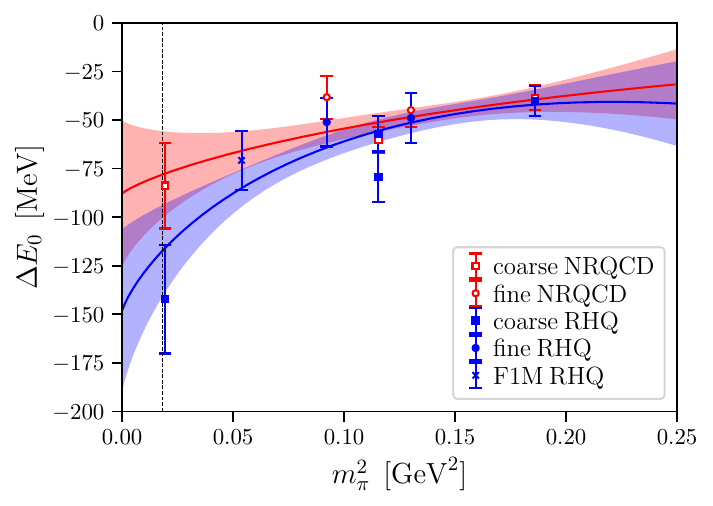}
\caption{\label{fig:chiralonly}Linear (left) and linear+log (right) chiral-only extrapolations of the NRQCD and RHQ results for the $T_{bb}$ binding energies obtained with the $3\times 3$ GEVP. The dashed vertical lines indicate the physical pion mass.}
\end{figure}

The chiral-only ``linear`` and ``linear+log'' fits are shown in Figure \ref{fig:chiralonly}; the ``linear+3/2'' fits look very similar to the ``linear+log'' fits and are not shown. The fit results for the coefficients of the $m_\pi^3$ or $m_\pi^2 \ln(m_\pi^2/\mu^2)$ terms are consistent with zero within $<1\sigma$, and the $\chi^2/{\rm d.o.f.}$ values are actually increased when adding these terms due to the reduction of d.o.f. by the extra parameter. Nevertheless, such terms are expected from chiral perturbation theory to be present, and the resulting fit curves show a possibly more physical behavior with a flattening of the pion-mass dependence in the higher-mass region. For the NRQCD data, we perform a model average over the three different chiral extrapolations with weights calculated using Eq.~(\ref{eq:modelweights}) to be approximately 56\%, 22\%, 22\%, respectively, giving $\langle \Delta E_0(m_{\pi,\text{phys}}) \rangle=(-74\pm17)$ MeV. We adopt the 10 MeV estimate of discretization and NRQCD systematic errors from Ref.~\cite{Leskovec:2019ioa}, yielding our final NRQCD-based prediction of
\begin{equation}
(m_{T_{bb}}-m_B-m_{B^*})_{\rm NRQCD}=(-74 \pm 17 \pm 10)\text{ MeV}.
\end{equation}

For the RHQ data, we focus on the chiral-continuum extrapolations, as these fits have better quality than the chiral-only extrapolations, and the continuum extrapolation should remove most discretization errors. In this case, we perform a model average over the fits in the last three rows of Table \ref{tab:extrapolations}, whose weights are found to be approximately 44\%, 27\%, 29\%. This yields our final RHQ-based prediction of
\begin{equation}
(m_{T_{bb}}-m_B-m_{B^*})_{\rm RHQ}=(-76 \pm 23)\text{ MeV}.
\end{equation}
Here, we expect that any higher-order discretization errors that remain after continuum extrapolation are negligible compared to the statistical uncertainty.

Finally, Fig.~\ref{fig:NRQCD3x3vs5x3} shows a comparison of our present NRQCD results for the binding energy, obtained with the $3\times3$ GEVP using local four-quark operators (demonstrated recently to be sufficient for the $T_{bb}$ \cite{Alexandrou:2024iwi}), with those of Ref.~\cite{Leskovec:2019ioa}, which were extracted from multi-exponential fits to correlation matrices including one or two scattering operators at the sink. A significant discrepancy is seen, which must be due to underestimated excited-state contamination, as the lattice parameters are identical. Due to the asymmetric correlation matrices used in Ref.~\cite{Leskovec:2019ioa}, the $\bar{b}\bar{b}ud$ energies obtained there may have a negative bias (resulting in an overestimate of the magnitude of the binding energy), while any bias due to insufficient $t_{\rm min}$ in the present re-analysis based on symmetric correlation matrices would be positive (resulting in an underestimate of the magnitude of the binding energy).

\begin{figure}
\includegraphics[width=0.49\linewidth]{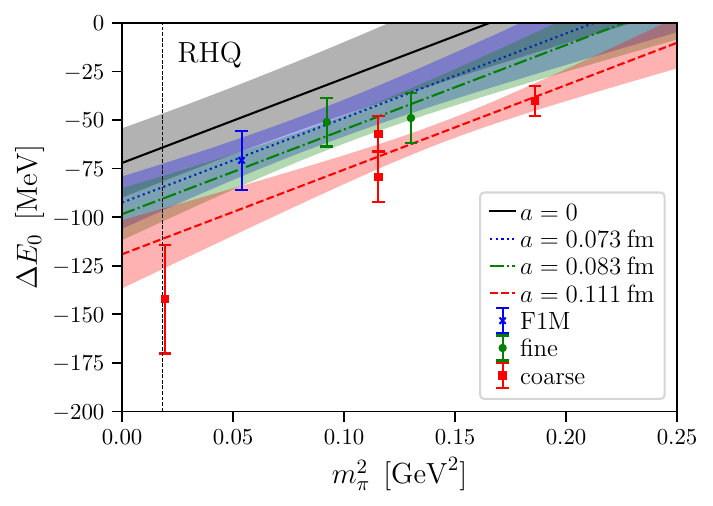} \includegraphics[width=0.49\linewidth]{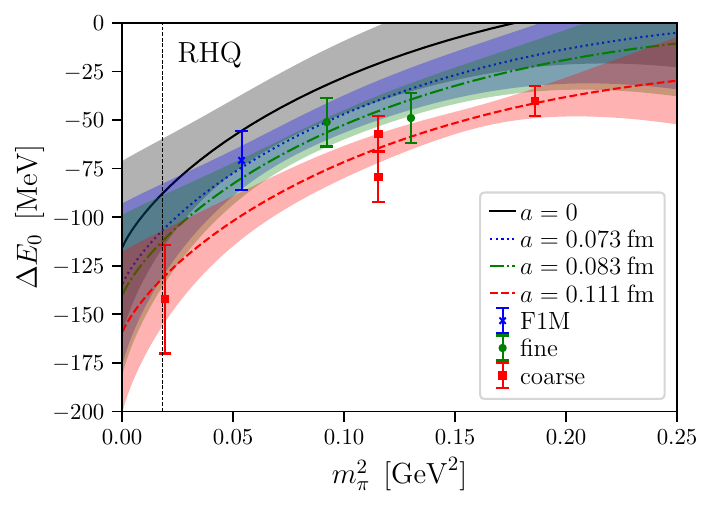}
\caption{\label{fig:RHQchiralcontinuum}Linear (left) and linear+log (right) chiral-continuum extrapolations of the $T_{bb}$ binding energies obtained with the RHQ action using the $3\times 3$ GEVP. The curves show the fit models evaluated for the three different lattice spacings for which we have data, and in the continuum limit.}
\end{figure}

\begin{figure}
\includegraphics[width=0.49\linewidth]{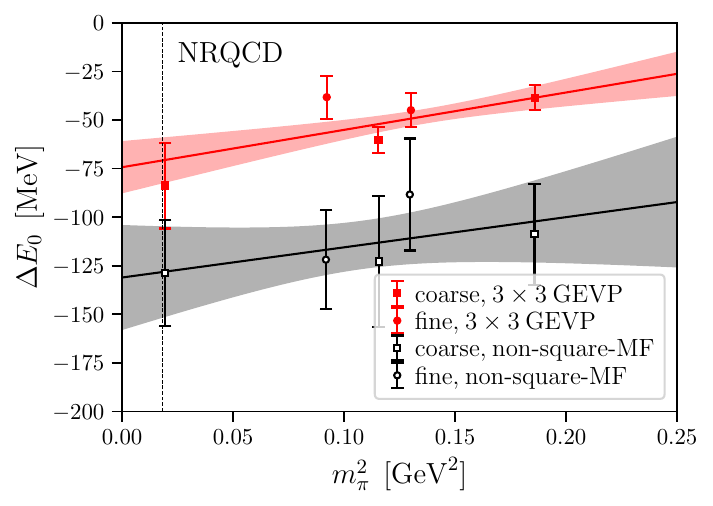}
\caption{\label{fig:NRQCD3x3vs5x3}Linear chiral-only extrapolations of NRQCD results for the $T_{bb}$ binding energy: comparison of the results obtained here with the GEVP from $3\times 3$ correlation matrices with only local four-quark operators with those obtained in Ref.~\cite{Leskovec:2019ioa} using multi-exponential matrix fits (MF) to correlation matrices including one or two scattering operators at the sink.}
\end{figure}

\FloatBarrier
\section{Conclusions}

This work provides (i) the first lattice-QCD determination of the $T_{bb}$ binding energy at the physical pion mass and in the continuum limit using an RHQ $b$-quark action, (ii) a controlled comparison of results obtained with NRQCD and RHQ $b$ quarks, and (iii) a comparison of results obtained with the GEVP from symmetric correlation matrices using local four-quark interpolating operators only with results from multi-exponential fits to non-square correlation matrices that additionally contain scattering operators at the sink.

Regarding (ii), we observe remarkably good agreement between the binding energies $\Delta E_0$ computed using the RHQ and NRQCD actions on the ensembles where we have data with both actions, especially at the heaviest pion masses where the statistical uncertainties are the smallest. Note that this finding is specific to our choices of matching coefficients and terms included in the actions. When comparing the chiral-only extrapolations in Fig.~\ref{fig:chiralonly}, we see a trend toward a larger $|\Delta E_0|$ at low pion mass for the RHQ case, but this appears to be mainly due to the addition of the C005LV ensemble and the doubled number of samples for C00078 in the RHQ case, and could be a statistical fluctuation. The good agreement between the RHQ and NRQCD results means that there is no evidence for large heavy-quark discretization errors.

Regarding (iii), as can be seen in Fig.~\ref{fig:NRQCD3x3vs5x3}, our reanalysis of the NRQCD data of Ref.~\cite{Leskovec:2019ioa} using only the symmetric $3\times3$ parts of the $5\times 3$ correlation matrices, and using the GEVP instead of multi-exponential matrix fits, gives substantially smaller $|\Delta E_0|$ (this was already observed in Ref.~\cite{Leskovec:2019ioa}, but the matrix fits with scattering operators at the sink were deemed more reliable at the time). As was recently demonstrated in Ref.~\cite{Alexandrou:2024iwi}, when including both local and scattering operators at both source and sink, a GEVP analysis of the full $5\times 5$ symmetric correlation matrices yields identical results for the ground-state energy as a GEVP analysis of the $3\times 3$ correlation matrices with local four-quark operators only\footnote{This finding is specific to the deeply bound $T_{bb}$ and we do not expect it to hold for less deeply bound systems.}. We therefore suspect that the results of Ref.~\cite{Leskovec:2019ioa} suffer from some negative bias in the $T_{bb}$ energies because the off-diagonal correlators with local four-quark operators at the source and scattering operators at the sink are susceptible to false plateaus (at intermediate times) below the true ground-state energy, which may distort even a multi-exponential fit. On the other hand, the present results using the  $3\times 3$ GEVP may be affected by some positive bias in the $T_{bb}$ ground-state energy, and hence underestimate $|\Delta E_0|$, because the principal correlator approaches the ground state from above. The model-averaging procedure employed here \cite{Jay:2020jkz} is designed to optimize the balance between statistical uncertainties and residual excited-state contamination, but it is conceivable that it still underestimates the total uncertainties due to the rapid growth of the noise with the Euclidean time separation.

\begin{figure}[h]
\includegraphics[width=0.6\linewidth]{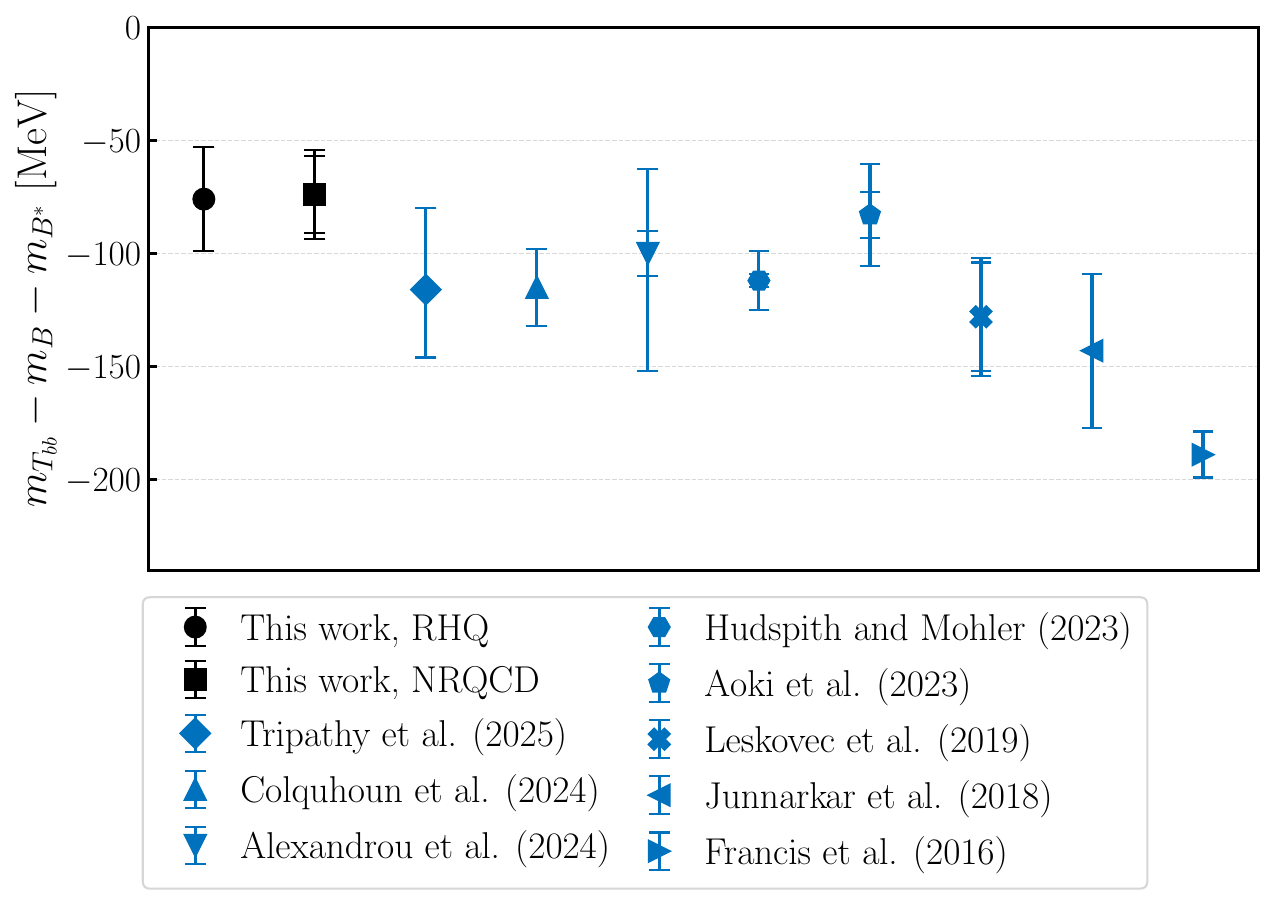}
\caption{\label{fig:comparison}Comparison of lattice-QCD predictions for the $T_{bb}$ binding energy. Only results extrapolated to the physical pion mass are shown \cite{Francis:2016hui,Junnarkar:2018twb,Leskovec:2019ioa,Aoki:2023nzp,Hudspith:2023loy,Alexandrou:2024iwi,Colquhoun:2024jzh,Tripathy:2025vao}, which were all obtained with NRQCD $b$ quarks (except for this work). The result of Ref.~\cite{Nagatsuka:2025szy} is omitted because that calculation did not include local-four-quark operators and likely missed the true ground state. The inner and outer error bars show the statistical and total uncertainties, where provided separately.}
\end{figure}

Our final results for the $T_{bb}$ binding energy are shown together with other lattice-QCD results in Fig.~\ref{fig:comparison}. We predict the smallest magnitude of the binding energy among these calculations. The particularly strong discrepancy with the result of Francis \textit{et al.} (2016) \cite{Francis:2016hui} is likely due to significant excited-state contamination with negative amplitudes due to the use of wall sources and point sinks, as also discussed and remedied by the authors in their follow-up work \cite{Hudspith:2020tdf,Colquhoun:2024jzh}. Our results are closer to more recent calculations. The excellent agreement between our NRQCD and RHQ lattice results further confirms the existence of a QCD-stable $T_{bb}$ tetraquark, with a smaller magnitude of the binding energy than suggested by the earliest lattice calculations.

\FloatBarrier
\section*{Acknowledgements}

We thank the RBC and UKQCD Collaborations for providing the lattice gauge configurations. We thank Luka Leskovec, Martin Pflaumer, and Marc Wagner for collaboration on Ref.~\cite{Leskovec:2019ioa}, whose data we re-analyzed here, and Will Detmold for sharing $\eta_b$ and $\Upsilon$ correlators computed with the RHQ action on two of the ensembles. We thank the developers of the Grid \cite{Boyle:2015tjk} and Grid Python Toolkit \cite{GPT} libraries, which we used to perform the new computations in this work. J.H. acknowledges support by the Deutsche 
Forschungsgemeinschaft (DFG, German Research Foundation) through the 
CRC-TR 211 ``Strong-interaction matter under extreme conditions'' -- 
project number 315477589-TRR 211. S.M.~is supported by the U.S. Department of Energy, Office of Science, Office of High Energy Physics under Award Number DE-SC0009913.
This research used resources of the National Energy Research Scientific Computing Center (NERSC), a Department of Energy User Facility using NERSC award HEP-ERCAP 31326.

\section*{Data Availability}

Machine-readable files containing, for each ensemble, the $B$, $B^*$, and $T_{bb}$ energies in lattice units, the $B^*-B$ hyperfine splitting in MeV, and the $T_{bb}$ binding energy in MeV, are provided in the Supplemental Material \cite{Supplemental}. The other data are available from the authors upon reasonable request.

\appendix

\section{Additional effective-energy plots}

\begin{figure}[htbp]
    \centering

    \includegraphics[width=0.45\textwidth]{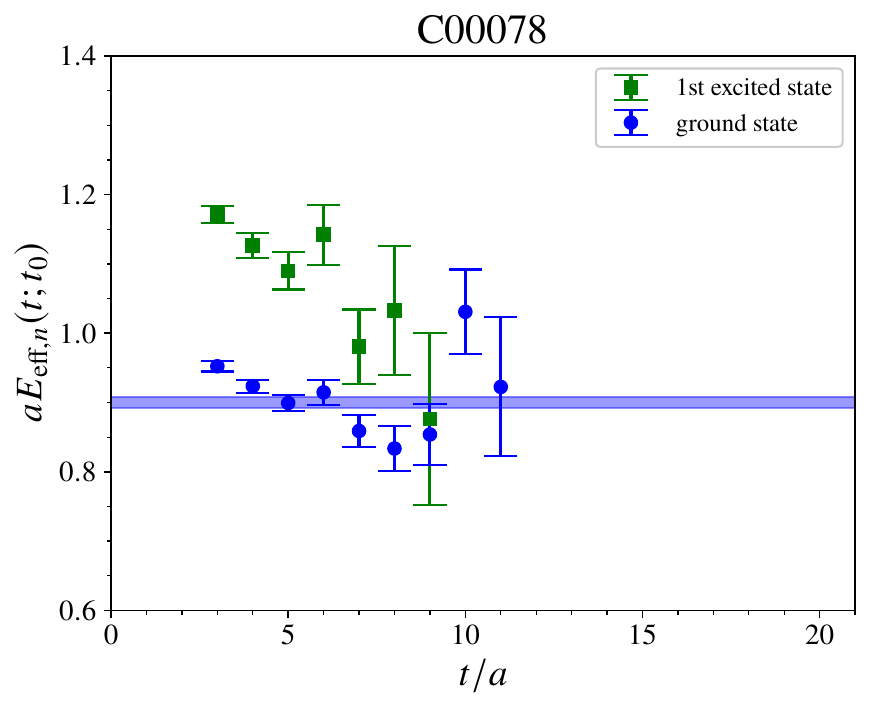}
    \hfill 
    \includegraphics[width=0.45\textwidth]{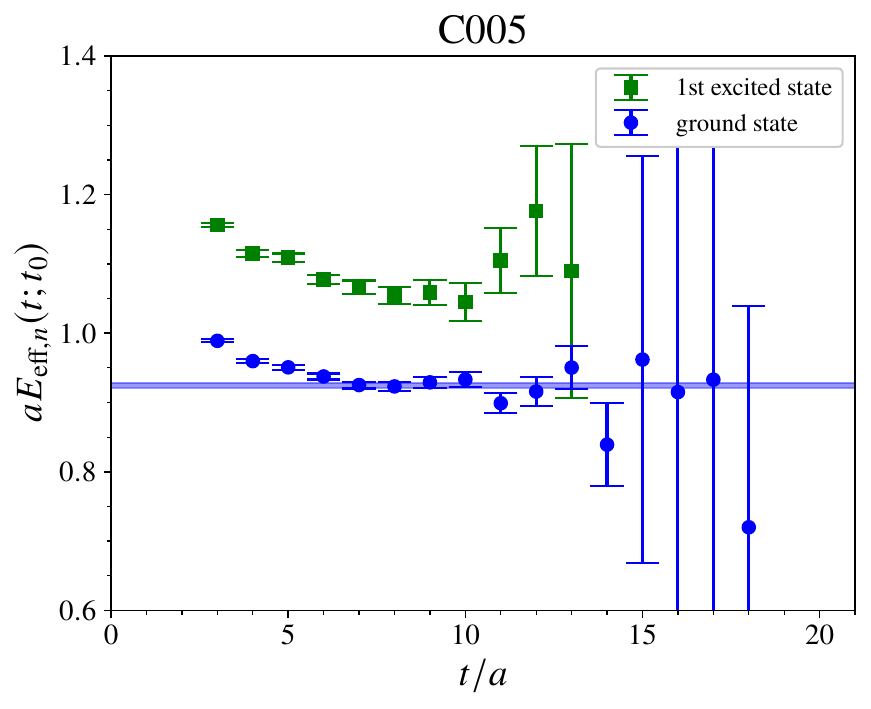}

    \medskip

     \includegraphics[width=0.45\textwidth]{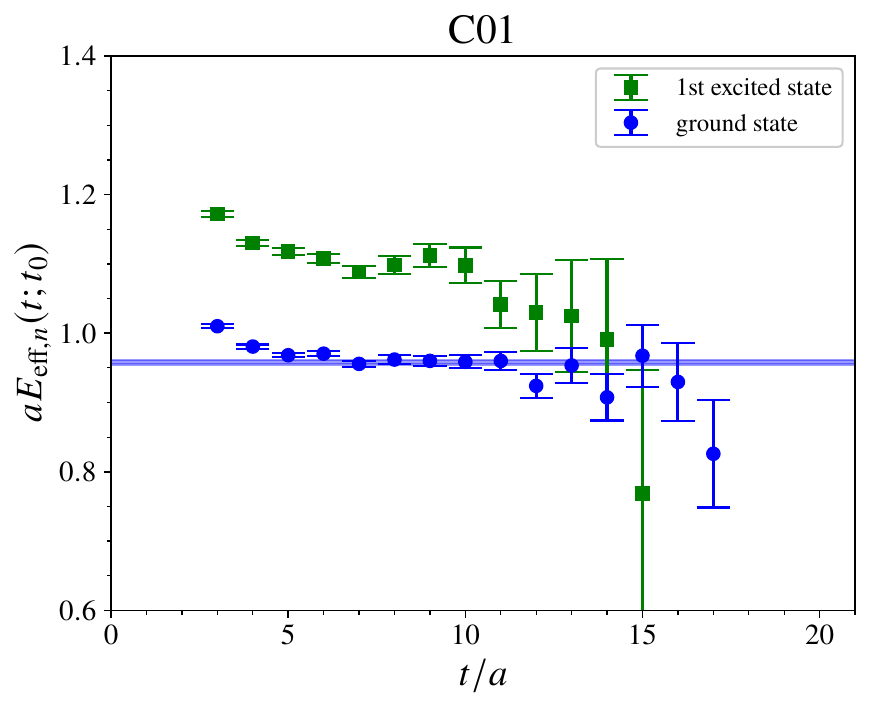}
    \hfill 
    \includegraphics[width=0.45\textwidth]{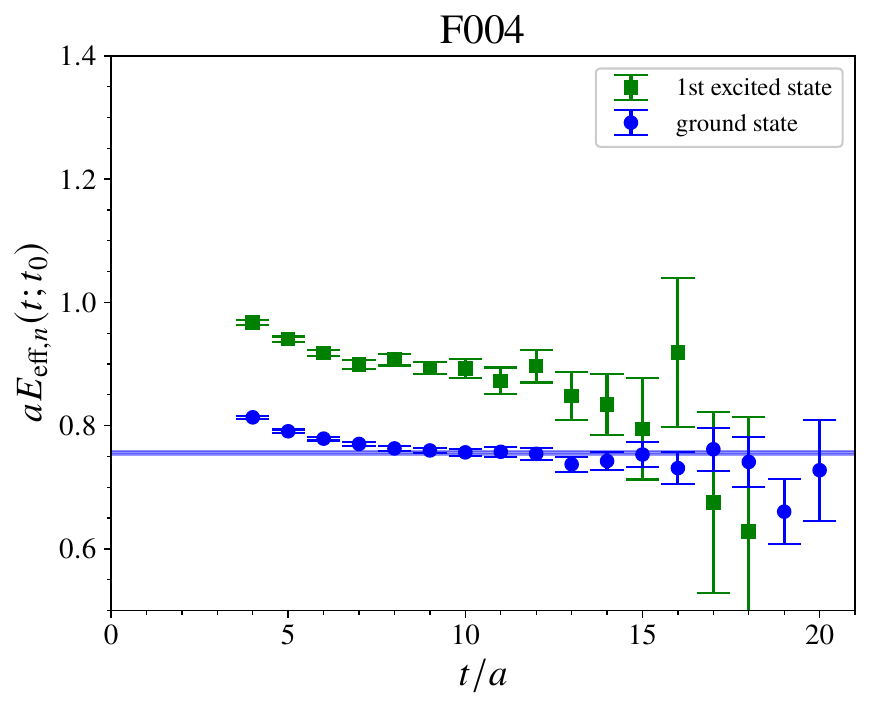}
\caption{\label{fig:EffNRQCD}Like Fig.~\protect\ref{fig:effective-energies-bbud}, but for the NRQCD data from the other ensembles.}
\end{figure}

\begin{figure}[h]
    \centering

    \includegraphics[width=0.45\textwidth]{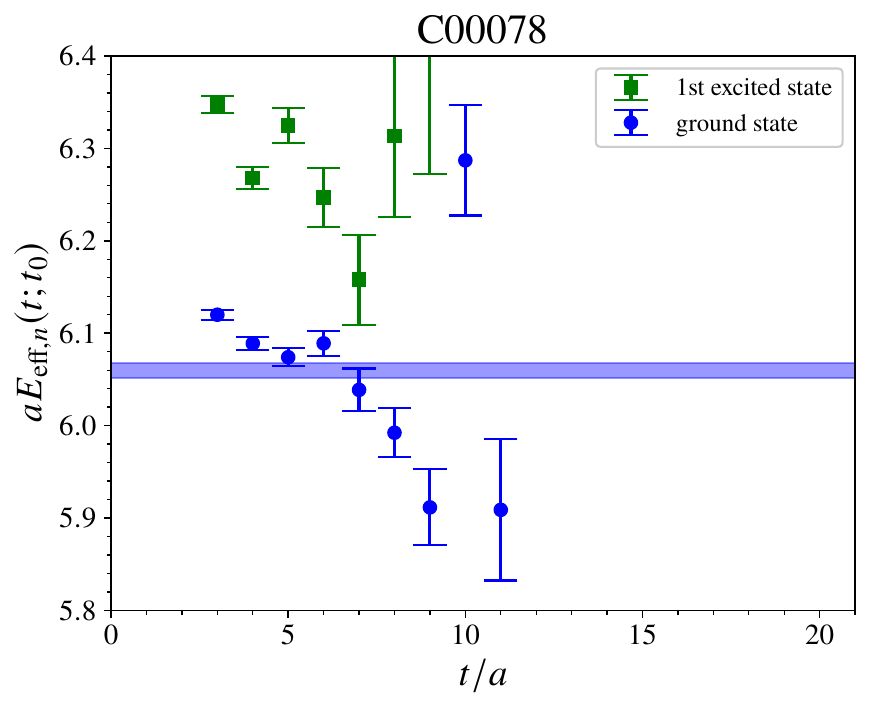}
    \hfill
    \includegraphics[width=0.45\textwidth]{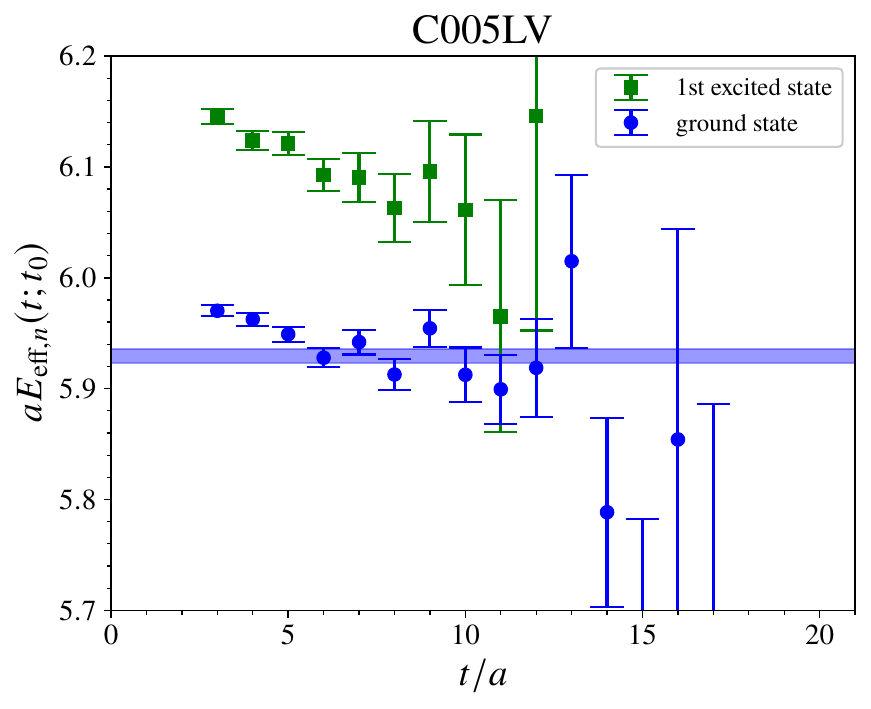}

    \medskip

    \includegraphics[width=0.45\textwidth]{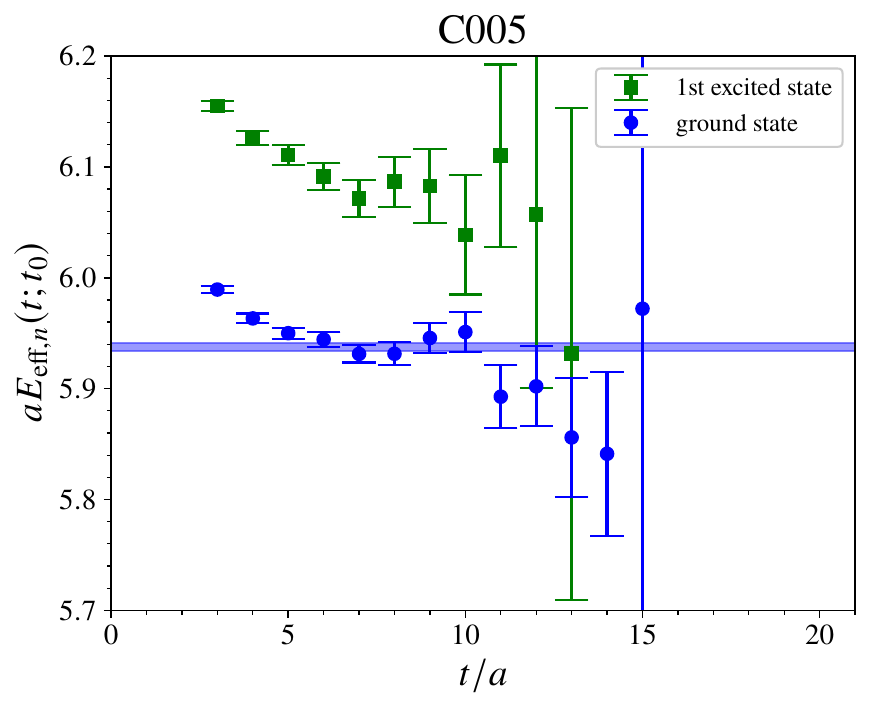}
    \hfill
    \includegraphics[width=0.45\textwidth]{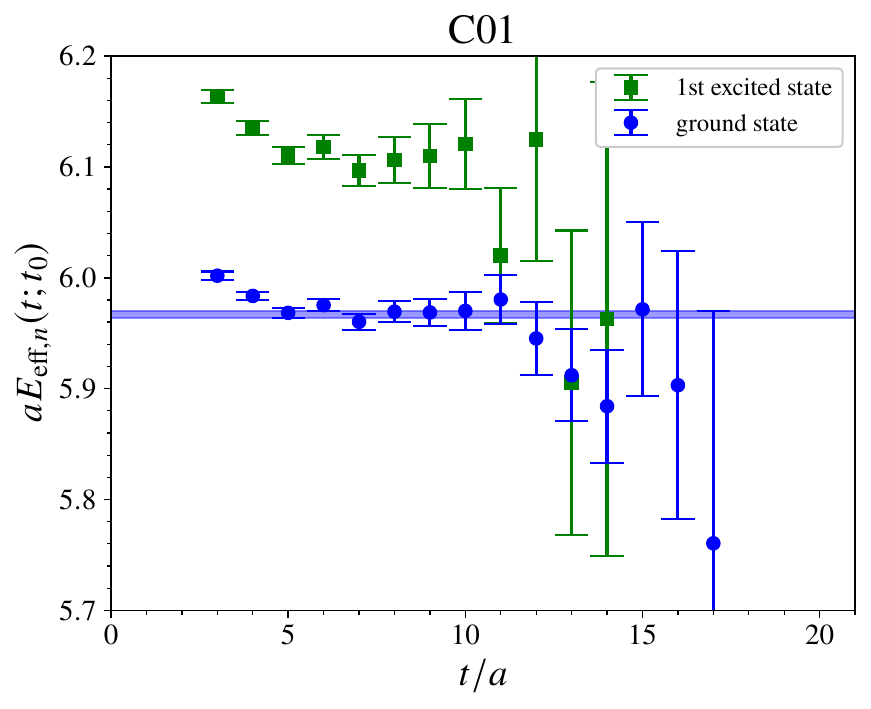}

    \medskip

    \includegraphics[width=0.45\textwidth]{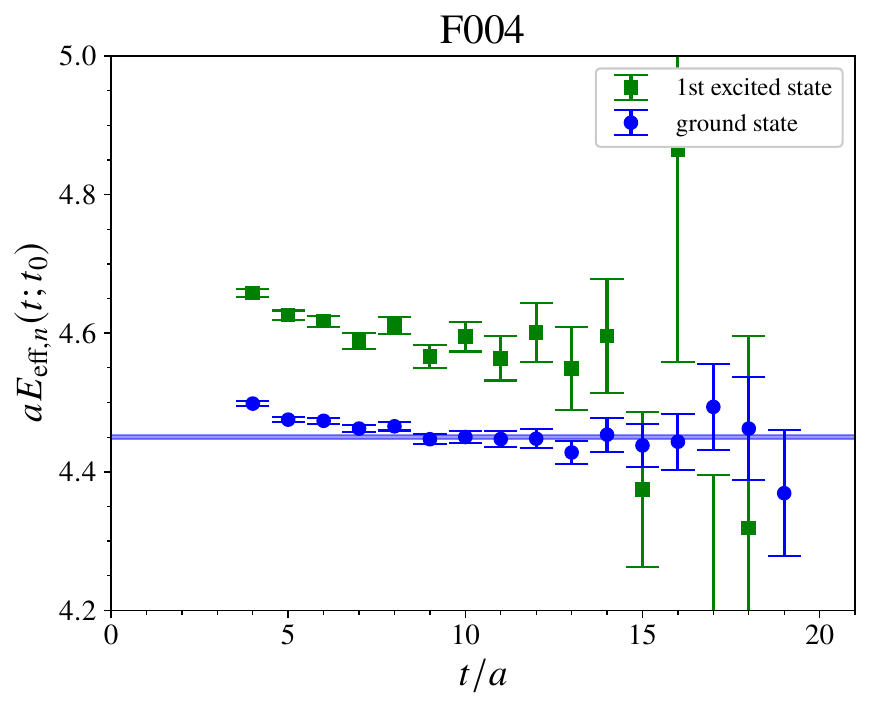}
    \hfill
     \includegraphics[width=0.45\textwidth]{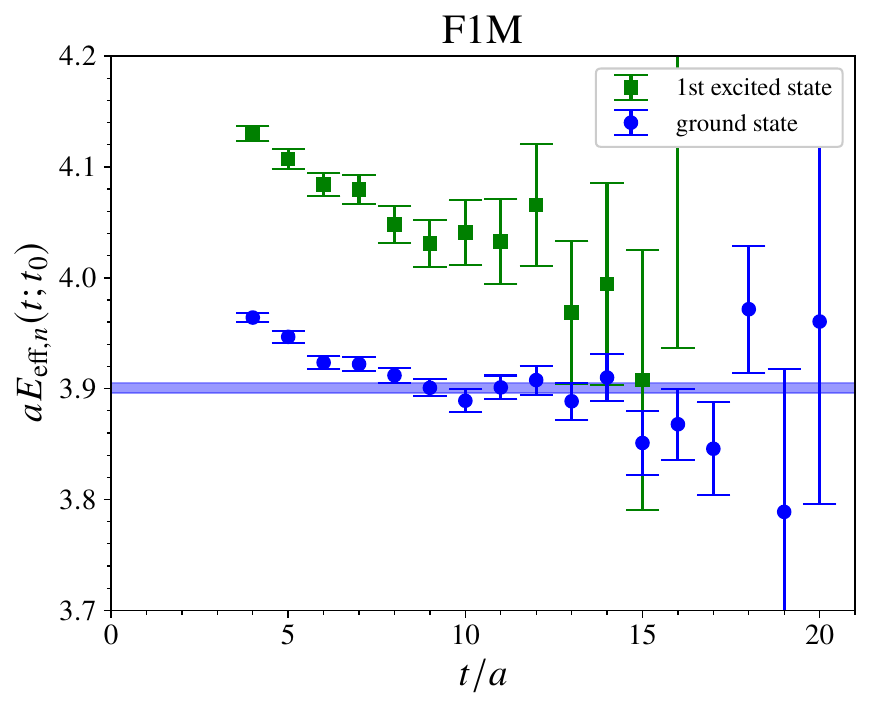}
    
\caption{\label{fig:EffRHQ}Like Fig.~\protect\ref{fig:effective-energies-bbud}, but for the RHQ data from the other ensembles.}
\end{figure}

Figures \ref{fig:EffNRQCD} and \ref{fig:EffRHQ} contain the $\bar{b}\bar{b}ud$ effective-energy plots for the ensembles not shown in the main text.

\FloatBarrier
\clearpage
\section{Additional model-averaging plots}

\begin{figure}[h]
    \centering
    
    \includegraphics[width=0.45\textwidth]{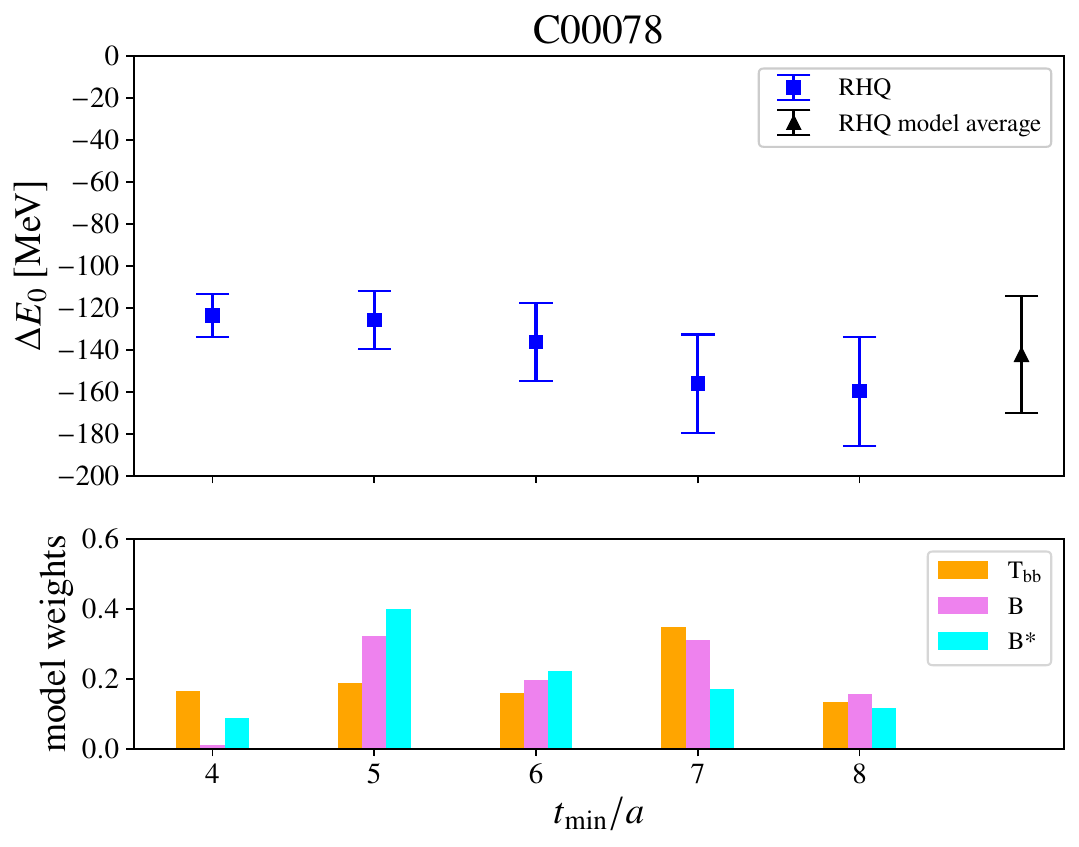}
    \hfill
    \includegraphics[width=0.45\textwidth]{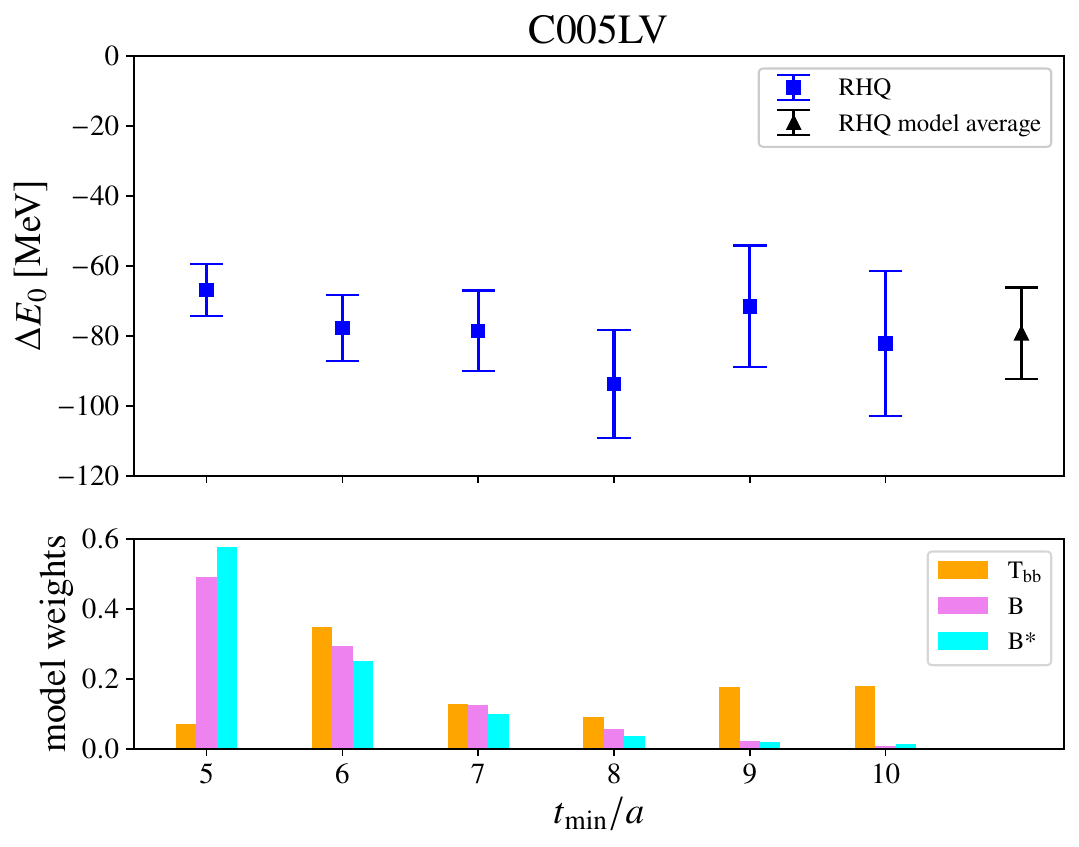}

    \medskip

    \includegraphics[width=0.45\textwidth]{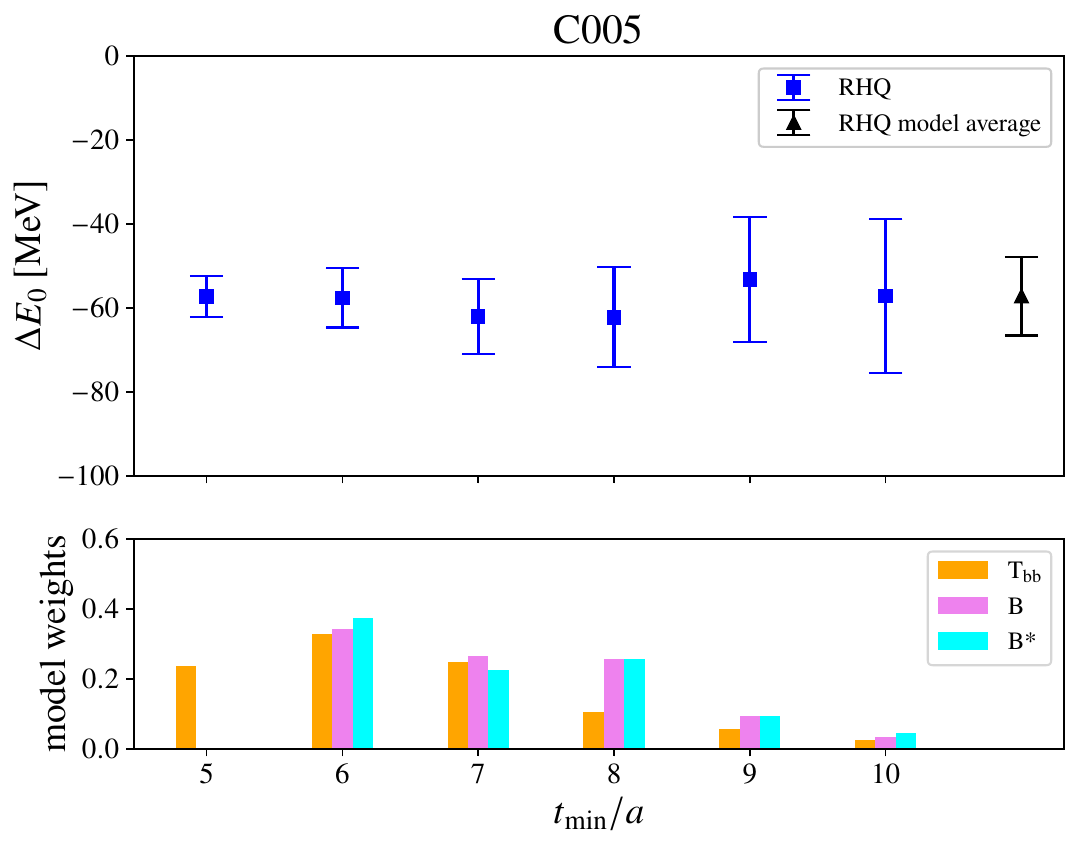}
    \hfill
    \includegraphics[width=0.45\textwidth]{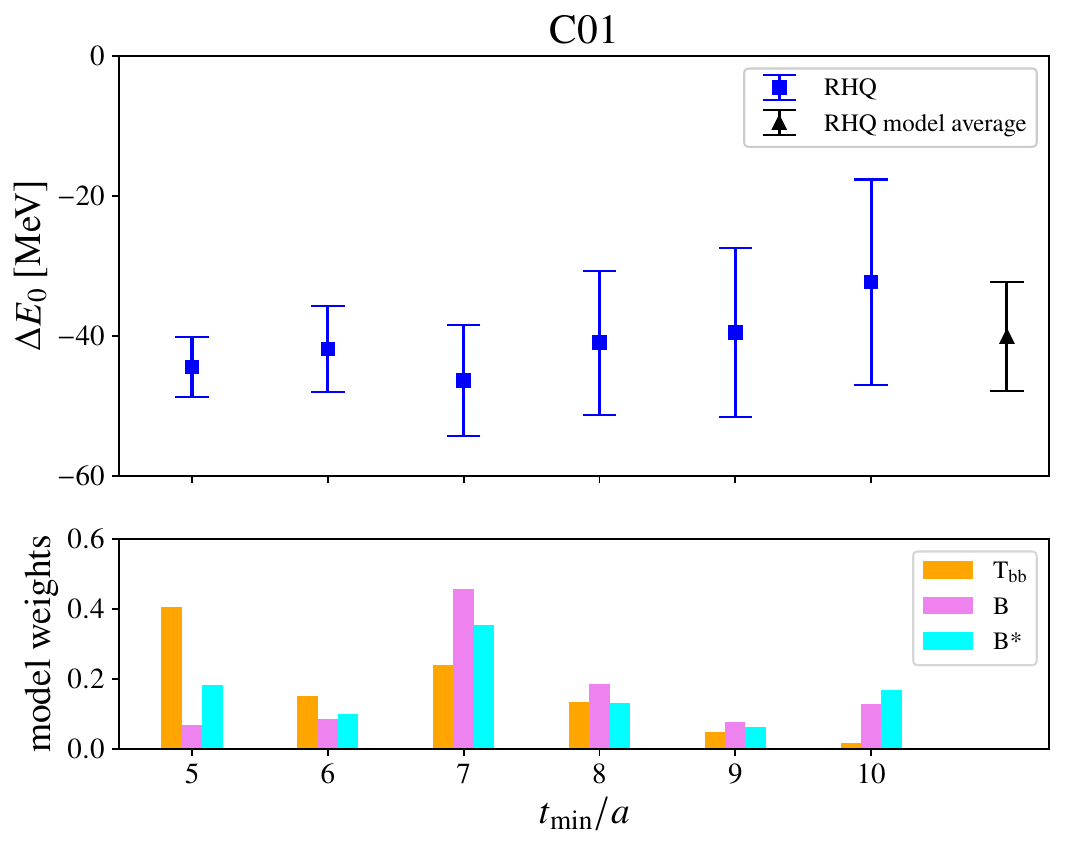}

    \medskip

    \includegraphics[width=0.45\textwidth]{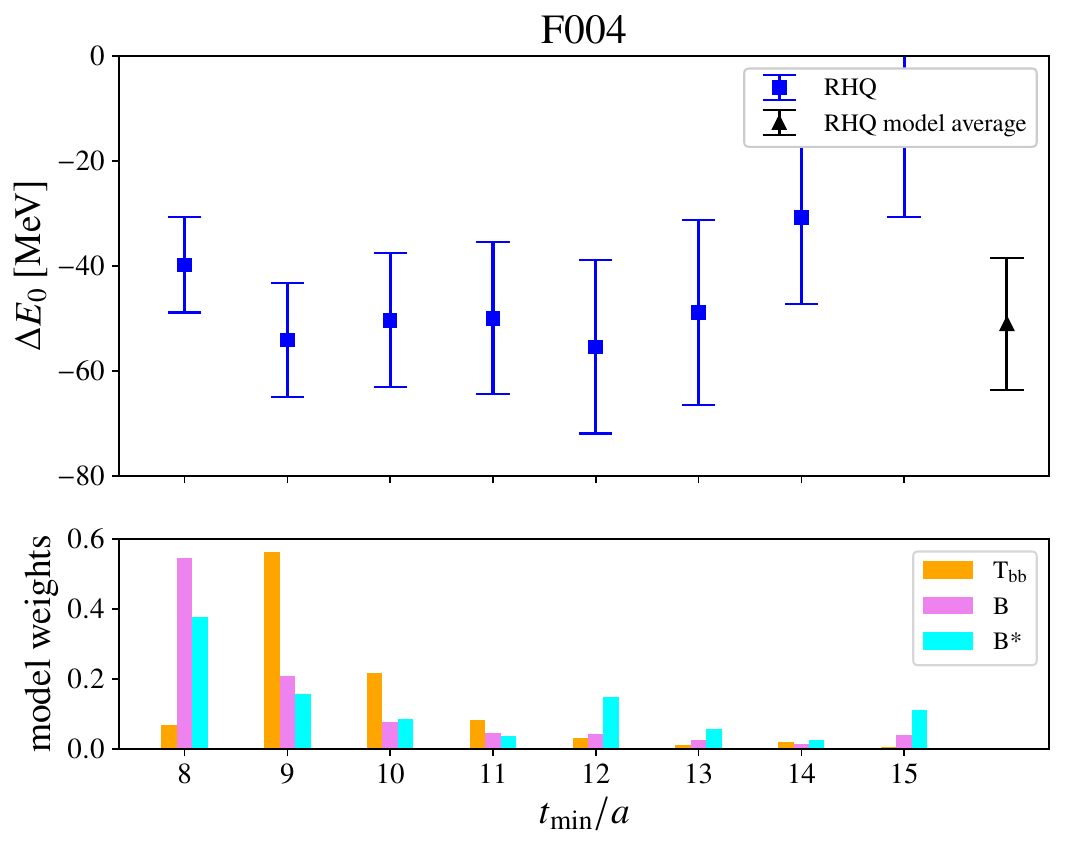}
    \hfill
     \includegraphics[width=0.45\textwidth]{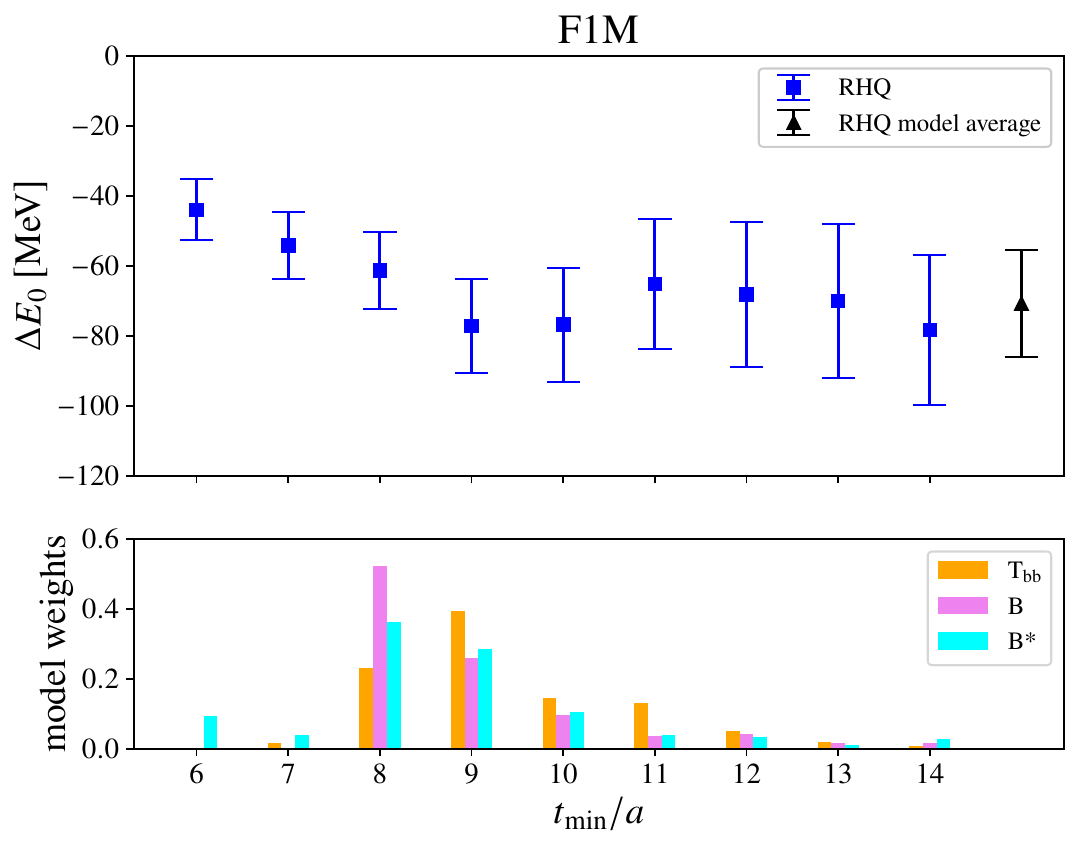}
    
\caption{\label{fig:DeltaEtminRHQ}Like Fig.~\protect\ref{fig:model-averaged-binding-energy}, but for the RHQ data from the other ensembles.}
\end{figure}

\begin{figure}[htbp]
    \centering

    \includegraphics[width=0.45\textwidth]{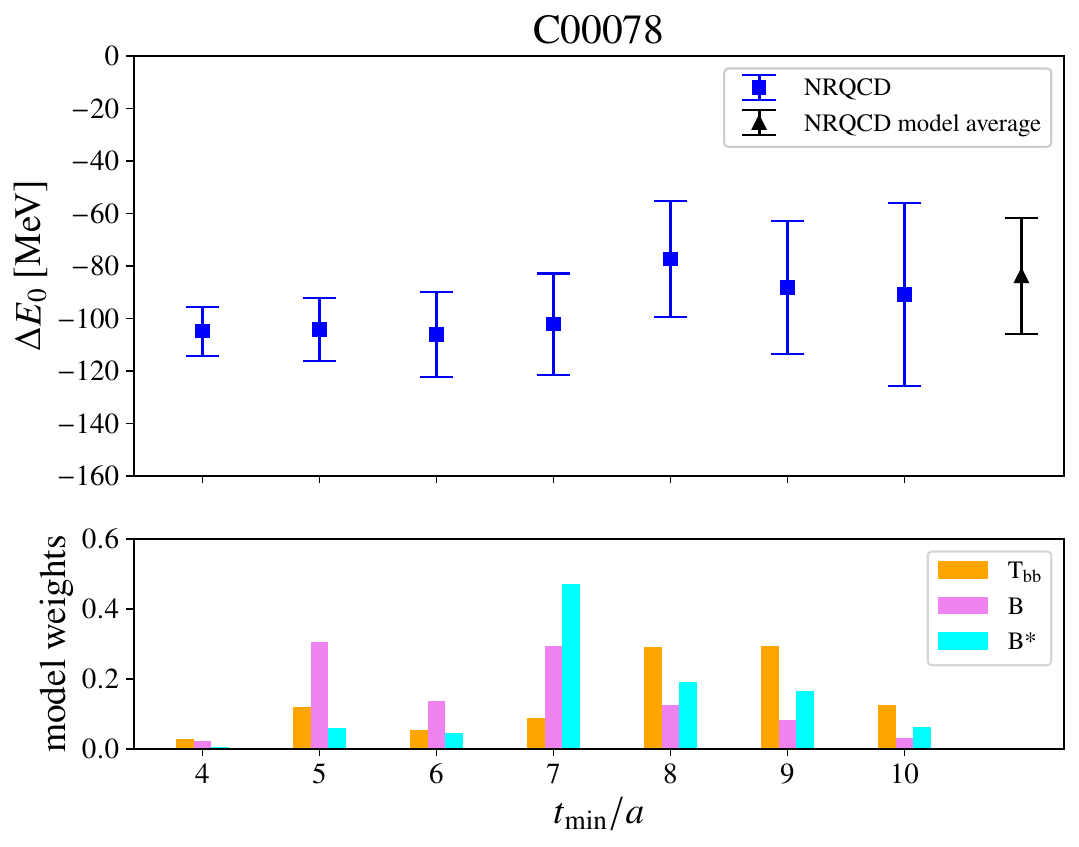}
    \hfill 
    \includegraphics[width=0.45\textwidth]{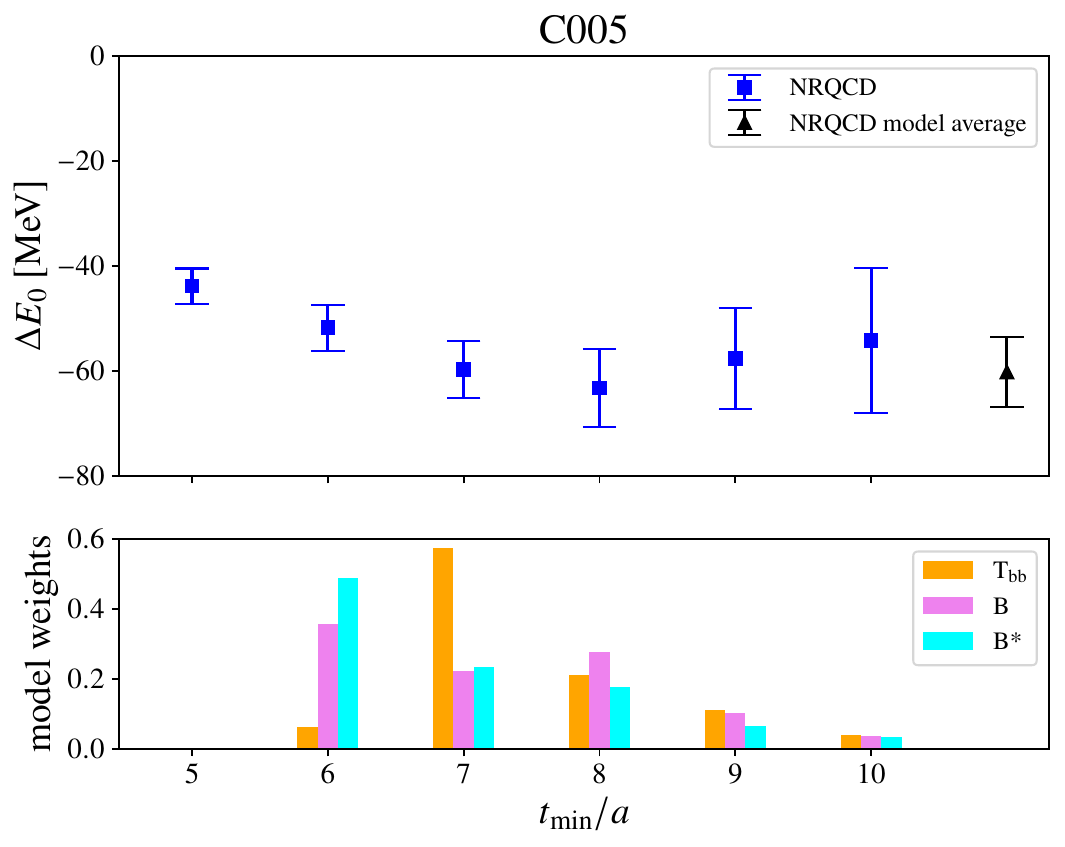}

    \medskip

     \includegraphics[width=0.45\textwidth]{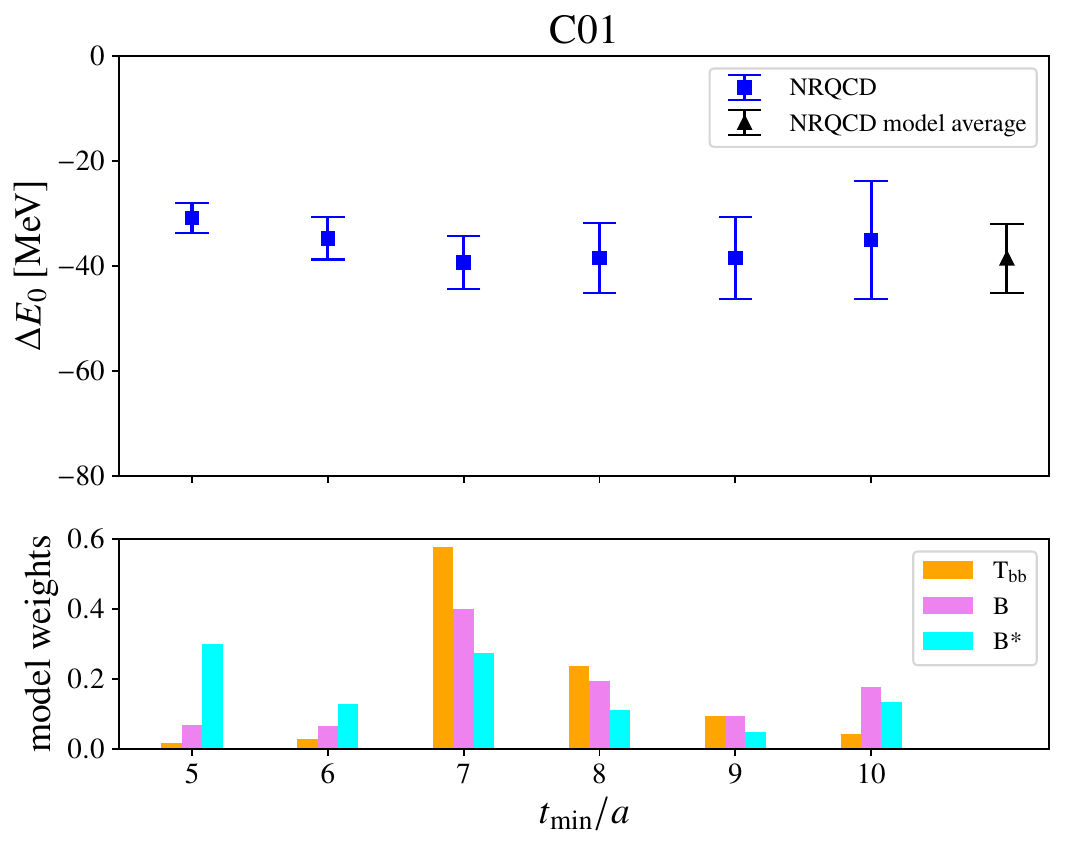}
    \hfill 
    \includegraphics[width=0.45\textwidth]{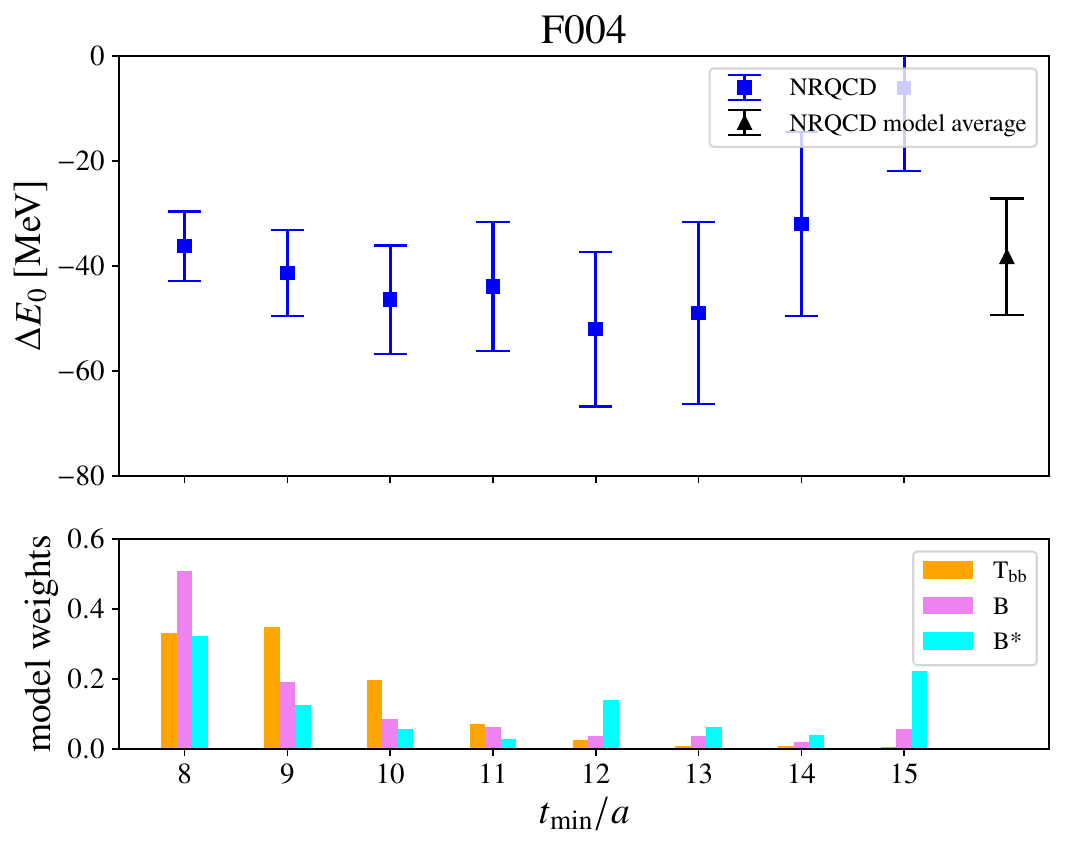}
\caption{\label{fig:DeltaEtminNRQCD}Like Fig.~\protect\ref{fig:model-averaged-binding-energy}, but for the NRQCD data from the other ensembles.}
\end{figure}

Figures \ref{fig:DeltaEtminRHQ} and \ref{fig:DeltaEtminNRQCD} contain the illustrations of the model averaging for the $T_{bb}$ binding energy for the
ensembles not shown in the main text.

\FloatBarrier

\section{Kinetic masses of heavy-light and heavy-heavy mesons}
\label{sec:kineticmasses}

In this section, we provide results for the kinetic masses of the $B$, $B^*$, $\eta_b$, and $\Upsilon$ mesons. These results are not used in the calculation of the tetraquark binding energies in this paper, and we do not have data for all ensembles, but they are of interest because they can be used to assess the accuracy of the $b$-quark mass tuning and the size of heavy-quark discretization errors. The rest and kinetic masses of a given hadron are defined by writing the square of its energy computed on the lattice with spatial momentum $\mathbf{p}$ as
\begin{equation}
E(\mathbf{p})^2 = m_{\rm rest}^2 + \frac{m_{\rm rest}}{m_{\rm kin}} \mathbf{p}^2 + ...,
\end{equation}
where the ellipses denote higher-order terms proportional to powers of the lattice spacing and higher-order polynomials of the components of $\mathbf{p}$ (restricted only by cubic symmetry) due to discretization errors. To extract the kinetic mass from numerical results for $E(\mathbf{0})=m_{\rm rest}$ and $E(\mathbf{p})$ at some small nonzero $\mathbf{p}$, two different definitions have been used in the literature (see, e.g., Refs.~\cite{Gray:2005ur,Meinel:2023wyg}), which become equivalent in the limit $\mathbf{p}\to \mathbf{0}$:
\begin{eqnarray}
\text{Definition 1: } m_{\rm kin}(\mathbf{p}) &=&  \frac{\mathbf{p}^2-[E(\mathbf{p})-E(\mathbf{0})]^2}{2[E(\mathbf{p}) - E(\mathbf{0})]}, \label{eq:mkin1} \\
\text{Definition 2: } m_{\rm kin}(\mathbf{p}) &=&  \frac{\mathbf{p}^2\,E(\mathbf{0})}{E(\mathbf{p})^2 - E(\mathbf{0})^2}. \label{eq:mkin2}
\end{eqnarray}
When using the RHQ action,  the ratio of rest to kinetic mass is also referred to as the square of the ``speed of light,''
\begin{equation}
c^2 = \frac{m_{\rm rest}}{m_{\rm kin}},
\end{equation}
and is desired to be close to 1.

When using the NRQCD action, $m_{\rm rest}$ does not include the contribution from the $b$-quark mass, and the above definition of $c^2$ is not helpful; instead, an alternative definition that describes the stability of $m_{\rm kin}$ computed with different values of $\mathbf{p}$ has been considered in the literature (see, e.g., Refs.~\cite{Gray:2005ur,Meinel:2010pv}).

In practice, the difference between the two kinetic-mass definitions (\ref{eq:mkin1}) and (\ref{eq:mkin2}) is usually very small and inconsequential when using the smallest possible momentum allowed by the periodic boundary conditions, $|\mathbf{p}|=\frac{2\pi}{L}$. In this section, we use definition (\ref{eq:mkin1}) with $|\mathbf{p}|=\frac{2\pi}{L}$ throughout. Our results for the kinetic masses of the $B$, $B^*$, $\eta_b$, and $\Upsilon$, for the NRQCD and RHQ actions, are shown in Tables \ref{tab:HLmkin} and \ref{tab:HHmkin}; Table \ref{tab:speedoflight} additionally gives the values of $c^2$ for the four different mesons computed with the RHQ action. Quark-disconnected diagrams were neglected in the calculation of the bottomonium correlation functions.

The mass parameter in the NRQCD action was tuned by demanding that the spin-averaged bottomonium kinetic mass agrees with experiment, but using the lattice spacings determined in Ref.~\cite{Meinel:2010pv}, which differ slightly from the values determined by RBC/UKQCD \cite{RBC:2014ntl,Boyle:2018knm} and used here. Therefore, the spin-averaged bottomonium kinetic masses converted to GeV using the lattice spacings determined by RBC/UKQCD  are close to, but differ slightly from, the experimental value. The spin-averaged heavy-light kinetic masses computed with the NRQCD action are also close to the experimental value.

For the RHQ action, whose three parameters were tuned using the $B_s$ rest mass, speed of light, and hyperfine splitting \cite{Meinel:2023wyg}, the $B$ and $B^*$ kinetic masses and speed of light are also found to be close to the correct values, as expected. On the other hand, the RHQ results for the bottomonium kinetic mass and speed of light show significant discretization errors, which are seen to diminish as the lattice spacing is reduced. This is also expected, as the typical heavy-quark momenta inside the $\eta_b$ and $\Upsilon$ are much higher than in the $B_s^{(*)}$, and the RHQ action can match only fewer terms in the heavy-quark expansion and does not have the same level of Symanzik improvement as the NRQCD action.

\begin{table}
\begin{tabular}{lccccccccccc}
\hline\hline
Label & $(a m_{B,{\rm kin}})_{\rm NRQCD}$ & $(a m_{B,{\rm kin}})_{\rm RHQ}$ & $(a m_{B^*,{\rm kin}})_{\rm NRQCD}$ & $(a m_{B^*,{\rm kin}})_{\rm RHQ}$ & $(m_{\rm kin,spinav})_{\rm NRQCD}$ [GeV] & $(m_{\rm kin,spinav})_{\rm RHQ}$ [GeV] \\
\hline
C00078 & 3.03(14) & 3.04(13) & 3.10(13) & 3.26(14) &  5.33(22) & 5.54(24) \\
C005LV & -- & 2.903(38) &  --  & 2.976(46) &  -- & 5.278(77)  \\
C005 & 3.002(40) & 2.972(39) & 2.993(42)  & 3.055(47) & 5.346(73) & 5.416(79)  \\
C01 & 3.034(38) &  --  & 3.030(40)  &  --  & 5.409(69) &  --  \\
F004 & 2.323(21) & 2.244(46) & 2.323(25)  & 2.283(54) & 5.536(57)  & 5.42(12) \\
F006 & 2.311(23) & 2.188(44) & 2.311(23)  & 2.245(53) & 5.513(54)  & 5.32(12) \\
F1M & -- & 1.882(49) & --  & 1.909(52) & -- & 5.15(13) \\
\hline\hline
\end{tabular}
\caption{\label{tab:HLmkin}Kinetic masses of the $B$ and $B^*$ mesons in lattice units and their spin averages $m_{\rm kin,spinav}=\frac{1}{4}m_{B,{\rm kin}}+\frac{3}{4}m_{B^*,{\rm kin}}$ in GeV. The experimental value of the spin-averaged mass is 5.31344(20) GeV \cite{ParticleDataGroup:2024cfk}. }
\end{table}

\begin{table}
\begin{tabular}{lccccccccccc}
\hline\hline
Label & $(a m_{\eta_b,{\rm kin}})_{\rm NRQCD}$ & $(a m_{\eta_b,{\rm kin}})_{\rm RHQ}$ & $(a m_{\Upsilon,{\rm kin}})_{\rm NRQCD}$ & $(a m_{\Upsilon,{\rm kin}})_{\rm RHQ}$ & $(m_{\rm kin,spinav})_{\rm NRQCD}$ [GeV] & $(m_{\rm kin,spinav})_{\rm RHQ}$ [GeV] \\
\hline 
C00078 & 5.402(13) & -- & 5.344(17) & -- & 9.267(34) & -- \\
C005LV & -- & -- & -- & -- & -- & -- \\
C005 & 5.418(11) & 7.1653(44) & 5.374(13) & 7.2861(71)  & 9.611(35) & 12.950(38) \\
C01 & 5.408(12) & -- & 5.371(13) & -- & 9.587(36) & -- \\
F004 & 4.0765(95) & 4.621(11)  & 4.055(12) & 4.651(14) & 9.663(45) & 11.067(51) \\
F006 & 4.0755(72) & -- & 4.0499(98) & -- & 9.667(41) & -- \\
F1M & -- & 3.781(13) & -- & 3.805(16) & -- & 10.286(38) \\
\hline\hline
\end{tabular}
\caption{\label{tab:HHmkin}Kinetic masses of the $\eta_b$ and $\Upsilon$ mesons in lattice units and their spin averages $m_{\rm kin,spinav}=\frac{1}{4}m_{\eta_b,{\rm kin}}+\frac{3}{4}m_{\Upsilon,{\rm kin}}$ in GeV. The experimental value of the spin-averaged mass is 9.4450(5) GeV  \cite{ParticleDataGroup:2024cfk}.}
\end{table}

\begin{table}
\begin{tabular}{lccccc}
\hline\hline
Label & $c^2_{B,{\rm RHQ}}$ & $c^2_{B^*,{\rm RHQ}}$ & $c^2_{\eta_b,{\rm RHQ}}$ & $c^2_{\Upsilon,{\rm RHQ}}$ \\
\hline 
C00078 & 1.006(45) & 0.948(43) & -- & -- \\
C005LV & 1.024(14) & 1.009(16) & -- & -- \\
C005 & 1.000(13) & 0.981(15) & 0.73057(45)   & 0.72127(70)   \\
C01 &  -- &  --  & -- & -- \\
F004 & 0.991(21) & 0.983(24)  & 0.8505(21)   &  0.8485(27)  \\
F006 & 1.019(21) & 1.002(24)  & -- & -- \\
F1M & 1.039(27) & 1.033(28) & 0.9163(32) & 0.9146(39)  \\
\hline\hline
\end{tabular}
\caption{\label{tab:speedoflight}The square of the ``speed of light'' obtained with the RHQ action for the $B$, $B^*$, $\eta_b$, and $\Upsilon$ mesons.}
\end{table}

\FloatBarrier
\providecommand{\href}[2]{#2}\begingroup\raggedright\endgroup

\end{document}